\documentclass[acmsmall]{acmart}

\usepackage{amsmath}
\usepackage{amsfonts}

\usepackage{amssymb}

\usepackage{graphicx}
\graphicspath{ {figures/} }
\usepackage{pgfplots}
\usetikzlibrary{pgfplots.groupplots}
\pgfplotsset{compat=1.18}
\usepackage{pgfplotstable}

\usepackage{caption}
\usepackage{subcaption}
\usepackage{float}

\usepackage{booktabs}
\usepackage{tabularx}
\usepackage{multirow}
\usepackage{array}
\newcolumntype{M}[1]{>{\centering\arraybackslash}m{#1}}
\newcolumntype{Y}{>{\centering\arraybackslash}X}

\usepackage{xurl}

\setcopyright{acmlicensed}
\acmJournal{PACMHCI}
\acmYear{2025} \acmVolume{9} \acmNumber{2} \acmArticle{CSCW069} \acmMonth{4}\acmDOI{10.1145/3710967}

\begin{document}

\title[An Analysis of Privacy Leakage from Real-World Facial Images on Twitter and Associated User Behaviors]{Everyone's Privacy Matters! An Analysis of Privacy Leakage from Real-World Facial Images on Twitter and Associated User Behaviors}

\author{Yuqi Niu}
\orcid{0009-0004-7624-4711}
\affiliation{%
  \institution{Shanghai Jiao Tong University}
  \city{Shanghai}
  \country{China}}
\email{niuyuqi@sjtu.edu.cn}

\author{Weidong Qiu}
\orcid{0000-0001-6428-1655}
\authornote{Corresponding authorsqiuwd@sjtu.edu.cn, S.J.Li@kent.ac.uk.}
\affiliation{%
  \institution{Shanghai Jiao Tong University}
  \city{Shanghai}
  \country{China}}
\email{qiuwd@sjtu.edu.cn}

\author{Peng Tang}
\orcid{0000-0001-6607-1280}
\affiliation{%
  \institution{Shanghai Jiao Tong University}
  \city{Shanghai}
  \country{China}}
\email{tangpeng@sjtu.edu.cn}

\author{Lifan Wang}
\orcid{0009-0001-4892-841X}
\affiliation{%
  \institution{Shanghai Jiao Tong University}
  \city{Shanghai}
  \country{China}}
\email{intefirm@sjtu.edu.cn}

\author{Shuo Chen}
\orcid{0009-0004-2131-7602}
\affiliation{%
  \institution{Shanghai Jiao Tong University}
  \city{Shanghai}
  \country{China}}
\email{csjssq@sjtu.edu.cn}

\author{Shujun Li}
\orcid{0000-0001-5628-7328}
\authornotemark[1]
\affiliation{%
  \institution{University of Kent}
  \city{Canterbury}
  \country{United Kingdom}}
\email{S.J.Li@kent.ac.uk}

\author{Nadin K\"{o}kciyan}
\orcid{0000-0002-2653-6669}
\affiliation{%
  \institution{University of Edinburgh}
  \city{Edinburgh}
  \country{United Kingdom}}
\email{nadin.kokciyan@ed.ac.uk}

\author{Ben Niu}
\orcid{0000-0003-2898-7495}
\affiliation{%
  \institution{Institute of Information Engineering, Chinese Academy of Sciences}
  \city{Beijing}
  \country{China}}
\email{niuben@iie.ac.cn}

\begin{abstract}
Online users often post facial images of themselves and other people on online social networks (OSNs) and other Web 2.0 platforms, which can lead to potential privacy leakage of people whose faces are included in such images. There is limited research on understanding face privacy in social media while considering user behavior. It is crucial to consider privacy of subjects and bystanders separately. This calls for the development of privacy-aware face detection classifiers that can distinguish between subjects and bystanders automatically. This paper introduces such a classifier trained on face-based features, which outperforms the two state-of-the-art methods with a significant margin (by 13.1\% and 3.1\% for OSN images, and by 17.9\%  and 5.9\% for non-OSN images). We developed a semi-automated framework for conducting a large-scale analysis of the face privacy problem by using our novel bystander-subject classifier. We collected 27,800 images, each including at least one face, shared by 6,423 Twitter users. We then applied our framework to analyze this dataset thoroughly. Our analysis reveals eight key findings of different aspects of Twitter users' real-world behaviors on face privacy, and we provide quantitative and qualitative results to better explain these findings. We share the practical implications of our study to empower online platforms and users in addressing the face privacy problem efficiently.
\end{abstract}



\received{January 2024}
\received[revised]{July 2024}
\received[accepted]{October 2024}

\maketitle

\section{Introduction}

People are increasingly sharing visual content (digital images and videos) on online platforms supporting user-generated content, especially online social networks (OSNs), driven by the rapid development of the Internet and the image-capturing capabilities of smartphones and other mobile devices. According to Statista~\cite{Statista2024OSNstatistics1, Statista2021OSN+UGC}, the number of OSN users reached 5.17 billion by July 2024. The increasing use of OSNs also has witnessed the fast increasing volume of online visual content (digital images and videos) especially those uploaded by OSN users, which has pushed image and video sharing portals YouTube, Instagram, and TikTok to be among the top five OSN platforms each with billions of active monthly users~\cite{Statista2024OSNstatistics2}.

Many images and videos shared online contain people's faces since human activities are at the core of our everyday lives. The facial information could be used to identify people in various contexts, and other more sensitive information (such as age and social relationships with others) could also be revealed from such facial information through inference~\cite{tkde-16-priguard}. Moreover, images can include faces of many others (e.g., bystanders) who may even not be aware that they are in the image. In such settings, the picture was taken and shared without the data subjects' explicit consent, and this can result in multi-user privacy conflicts~(MPCs) and also the violation of data protection laws such as the EU and the UK's GDPR~\cite{gdpr}. The privacy issues can lead to other online harms, such as cyberbullying, identity theft, and re-identification in the real world. 

The existing work shows that people have concerns about unauthorized appearance of their faces in images posted online~\cite{Rashidi2018privacy_management}. Researchers have proposed different methods~\cite{bo2014privacy,shu2018cardea, li2018politecamera, zhang2018cloak} to address the MPCs in the context of sharing photos. Nevertheless, such privacy protection methods can usually only protect users who have set privacy preferences within specific systems and image co-owners who actively participate in the shooting activity, but fail to work for bystanders since they may be unacquainted with the photographer and the main subject(s), and even do not perceive the shooting activities. Some prevention mechanisms have been proposed by the multi-agent systems community~\cite{toit-17-priarg, toit-17-prinego}; however, there is no automated analysis of images being shared by users who may have different privacy needs.

Recently, the privacy of bystanders has gained more attention. According to a survey conducted in 2016~\cite{aditya2016pic}, more than 95\% of the participants believed that the privacy needs of bystanders should be taken into account. However, there is not enough research to understand and address such needs. We identified two main research gaps. \textbf{First, bystander privacy in images is an understudied area of research.} Our literature review led to only two past studies, conducted by Hasan et al.~\cite{hasan2020bystander_privacy} and Darling et al.~\cite{Darling2019, DarlingLL20}, respectively, where they developed machine learning-based classifiers to automatically detect bystanders within an image to support necessary protection. However, Hasan et al.'s method~\cite{hasan2020bystander_privacy} requires the whole-body view of people and, therefore, cannot be applied to many online images where online users often tend to share close-up pictures without the whole-body view. While Darling et al.~\cite{Darling2019, DarlingLL20} did use features based on face regions, addressing the issue of relying on the full-body view, the features they used are not sufficient. Their approach lacks a comparison and correlation between the local features of the face region and the overall features of the photo. \textbf{Second, none of past related work has provided insights regarding real-world behaviors of online users about face privacy.} In this paper, we aim to fill these two important research gaps.

We first introduce \textit{a novel machine learning-based bystander-subject classifier}, which is trained on mainly face-based features that are more readily available in online images (such as those shared on OSNs). To support the development (training, validation, testing, and performance comparison) of the new classifier, we constructed three new labeled datasets. Our new classifier achieved an accuracy of above 93\% on all datasets, and also significantly outperformed the state-of-the-art classifiers proposed by Hasan et al.~\cite{hasan2020bystander_privacy} and Darling et al.~\cite{Darling2019,DarlingLL20} with a large margin: an accuracy of 95.8\% vs 82.7\%~\cite{hasan2020bystander_privacy} (by 13.1\%) vs 92.7\%~\cite{Darling2019,DarlingLL20} (by 3.1\%) for OSN images, and an accuracy of 93.2\% vs 75.3\%~\cite{hasan2020bystander_privacy} (by 17.9\%) vs 87.3\%~\cite{Darling2019,DarlingLL20} (by 5.9\%) for non-OSN images. Second, we conduct a large-scale validation of our bystander-subject classifier via \textit{the development of a semi-automated framework} for quantitative and qualitative analysis of the face privacy problem on OSNs. For this, we collected 27,800 real-world images containing at least one face, posted publicly by 6,423 Twitter\footnote{Since we collected our data, the platform has been renamed to X. In this paper, we still use the old name Twitter.} users. Third, we \textit{analyze this dataset to understand online users' behaviors in the real world}. Our analysis of the images led to eight key findings supported by quantitative and qualitative results, providing new evidence on different aspects of Twitter users' behaviors around face privacy. The findings cover general user behaviors and facts about how Twitter users posted images containing faces, the lack of face-protecting behaviors of most users, behaviors of a minority of users who chose to protect some faces, and potential leakage of social attributes of people whose faces are included in such images. Different than previous work on face privacy through empirical studies such as user surveys~\cite{Photoconflicts_such2017, photosharing_Amon2020}, our analysis is based on real-world data and at a much larger scale. Our findings have practical implications for online users by raising privacy awareness and also for platforms by assisting them in the development of privacy protection tools.

Our key contributions can be summarized as follows\footnote{The source code and data used in this paper, together with the relevant instructions for reproducing our results, are available at \url{https://github.com/Yuqi-Niu/Bystander-Detection}.}:
\begin{itemize}
\item We designed a new machine learning-based bystander-subject classifier with face-based features, which is more applicable for analyzing online social media images and was able to outperform the most recent state-of-the-art solutions significantly~\cite{hasan2020bystander_privacy, Darling2019, DarlingLL20}.

\item Based on our new bystander-subject classifier, we developed a semi-automated framework for quantitative and qualitative analysis of the face privacy problem on OSNs. This framework was applied to a large dataset of 27,800 Twitter images including at least one face, showing that our bystander-subject classifier worked well in a real-world setting.

\item Based on the results of the 27,800 Twitter images, for the first time in the literature, we conducted a large-scale quantitative and qualitative analysis of the face privacy problem, leading to 8 key findings covering different aspects of the face privacy problem on Twitter, which provides important insights on online users' behaviors and how to protect people's privacy online more effectively.
\end{itemize}

The rest of this paper is organized as follows. Section~\ref{sec:related_work} discusses related work in the domain of image privacy. In Section~\ref{sec:overview}, we explain the context of our work and our overall methodology. Section~\ref{sec:sub_and_by} describes the detailed design and performance evaluation of our new bystander-subject classifier. Section~\ref{sec:pipeline} introduces the semi-automated framework for analysis of the face privacy problem. Section~\ref{sec:privacy} reports results of our large-scale analysis of the 27,800 images collected from Twitter. Section~\ref{sec:discussions} presents limitations and our future directions. We discuss our research ethics in Section~\ref{sec:Ethical_Considerations} and we conclude with Section~\ref{sec:conclusion}.

\section{Related Work}
\label{sec:related_work}

In this section, we discuss three main areas in face privacy research: detection of bystanders in images, protection of face privacy, and analysis of privacy settings of OSN platforms.

\subsection{Automatic Bystander Detection in Images}
\label{subsec:bystander_detection}

Researchers have studied bystander privacy (i.e., persons who are not device owners or controllers) in virtual, audio, and mixed reality~\cite{Corbettmxbystander, OHaganSGMMKM22, DeldariFPY23} as well as mobile live streaming~\cite{WuGWL23}, and proposed methods to identify bystanders in these contexts~\cite{CorbettDSHJ23}. However, there is limited research on automatic bystander detection in images. Li et al.~\cite{li2019hideme} proposed leveraging an image’s metadata to calculate shooting distance for identifying bystanders. Yet, the variability in shooting scenarios and photographers’ habits make it difficult to establish a consistent decision threshold based solely on distance. In addition, data such as the focal length of the lens required to calculate the shooting distance can be difficult to obtain in many cases.

Hasan et al.~\cite{hasan2020bystander_privacy} looked at automated bystander detection in photos. They first extracted some proxy features including human-related features extracted by ResNet50~\cite{HeZRS16}, body-pose related features extracted by OpenPose~\cite{CaoHSWS21}, and emotional features estimated from facial expressions. Then, they trained three models to predict three high-level concepts (pose, replaceable, and photographer's intention) as new features. The final features their bystander classifier uses include the human body size and the three high-level concepts. Although their classifier achieved reasonable performance on their dataset with non-OSN images, their work has the following issues: 1) they did not test it on real-world OSN images; 2) their classifier unnecessarily requires the presence of the whole human body in the image, which can limit the generalizability of their classifier to many real OSN images; and 3) their definition of the concept of bystanders is limited to reflect privacy-related aspects (we will further discuss this point in Section~\ref{subsec:bystander_defination} with greater detail). Darling et al.~\cite{Darling2019} proposed a facial feature-based bystander classifier that utilizes face size, head pose, blur level, and gaze vector extracted from the face region as features to train a machine learning model. In their subsequent work~\cite{DarlingLL20}, they compared this feature-based method with a CNN-based approach that uses the face area of the input image directly as input. Their results indicated that the feature-based solution achieved higher accuracy. However, their work had several limitations, including not being tested on real-world OSN images and the lack of a clear definition of bystanders in relation to privacy concerns.

While there are numerous datasets~\cite{moschoglou2017agedb, cao2018vggface2, maze2018iarpa} used for face recognition and other face-related work, to the best of our knowledge, only two datasets have been developed specifically for automatic bystander detection. One such dataset was recently reported by Hasan et al.~\cite{hasan2020bystander_privacy} in their S\&P '20 paper. They cropped 5,000 images from 2,583 images with at least one person, which were selected from the Google Open Image Dataset V4~\cite{kuznetsova2020open}. They used an online survey to label the person in each of the 5,000 images as a subject or a bystander, by asking recruited human participants to view the 2,583 images. The images unfortunately do not represent typical images posted on OSNs, which often do not have the whole human body. Darling et al.~\cite{DarlingLL20} released another dataset containing 515 faces cropped from 222 photos sourced from social platforms, public news sites, and image repository sites. This dataset, however, has several limitations, e.g., it is relatively small in scale, and only the cropped 515 face images -- not the 222 original images -- are publicly available.

Our work addresses the limitations of the above-reviewed work on automatic bystander detection by providing a more effective machine learning-based classifier and three new datasets.

\subsection{Image Privacy Protection Solutions}

Various methods have been proposed to address privacy issues caused by unauthorized image capturing and online photo sharing. One class of methods includes disabling camera sensors by near-infrared pulsating lights~\cite{truong2005preventing}, using broadcast commands~\cite{tiscareno2014systems} and pre-compiled contextual rules~\cite{kapadia2007virtual, jung2014courteous, steil2019privaceye}. However, such methods cause inconvenience to photographers. Some researchers~\cite{sharif2016accessorize, yamada2012use} proposed that people take proactive measures to prevent inference of their identity by wearing hardware devices that can hide or interfere with identifiable features (such as facial features), but such hardware devices can introduce discomfort and additional costs to users. Similar drawbacks were also reported in the context of wearable glasses~\cite{Krombholz2015glass}. Some other researchers~\cite{dimiccoli2018mitigating, alharbi2019mask} proposed to automatically blur the whole or part of the image to improve people's willingness to be captured. However, such coarse-grained methods often fall short of meeting diverse privacy needs of different people in different shooting scenarios, especially when multiple people of different groups (e.g., main subjects and bystanders) are photographed for a single photo.

To avoid MPCs and achieve finer-grained privacy protection, Kandappu et al.~\cite{KandappuSX21} analyzed multiple life log images to identify privacy-sensitive factors and then used blurring technology to selectively blur parts of the photos, thereby alleviating privacy issues and achieving a balance between privacy and usability. Some researchers studied the use of tags~\cite{pallas2014offlinetags}, QR codes~\cite{bo2014privacy}, and gestures~\cite{shu2017your} for communicating people's privacy preferences to photographers. However, malicious actors can also infer privacy preferences communicated using such marks, therefore introducing a new vector of privacy leakage (e.g., a malicious actor then takes an image of the individual communicating their privacy preference). Some other related work~\cite{toubiana2012photo, shu2018cardea, aditya2016pic, hu2012multiparty, xu2015my, ilia2015face, xu2018trust} associated facial information with access control policies to achieve image- or data-level protection by collaborative management of shared data. PrivacyCamera~\cite{li2016privacycamera} and PoliteCamera~\cite{li2018politecamera} are two example solutions adopting the collaborative scheme but overlook the security of information transmitted. Such approaches assume a trusted third party defining and enforcing such policies, which does not match many real-world scenarios of image privacy. Zhang et al.~\cite{zhang2018cloak} proposed a graph-matching scheme and a vector distance protocol to solve the above issues, but people have to register with the scheme to be able to formulate privacy protection strategies.

Zheng et al.~\cite{Zheng2022non-consensual_photo} extracted gaze and head direction features to train a neural network model to identify ``unaware parties'' in photos. Their work is related to the bystander research discussed in the previous section, but they considered ``unaware parties'' and bystanders (the definition of Hasan et al's ~\cite{hasan2020bystander_privacy}) two different concepts in their study.

In real-world applications, OSN platforms currently have considered only addressing privacy conflicts between the image uploader and the uploader's friends appearing in the same image, while ignoring people who are not connected with the uploader (e.g., bystanders who do not have an account).

\subsection{Privacy-related Analysis on OSNs}

A rich body of previous work looked at measuring and analyzing various aspects related to privacy on OSNs. For instance, Hassan et al.~\cite{hassan2018analysis} performed a systematic analysis of privacy behaviors and threats in fitness tracking OSNs. Kandappu et al.~\cite{KandappuSX21} analyzed how a life-logging service provider could glean sensitive information by correlating life-logs uploaded by several life-loggers. Halimi et al.~\cite{Halimi2021OSNprivacy} proposed a machine learning model to infer the vulnerabilities of OSN users for profile-matching privacy risks with high accuracy by only analyzing publicly available information of their local profiles in a targeted anonymous OSN. 

Some past work focused on OSN privacy settings. Liu et al.~\cite{liu2011analyzing} measured the disparity between the desired and actual privacy settings and quantified the magnitude of Facebook's privacy management problems. Mondal et al.~\cite{mondal2019moving} studied the privacy settings of Facebook posts and developed a tool to infer potentially mismatched privacy settings. Reichel et al.~\cite{reichel2020have} studied how to tailor OSN privacy settings to users in resourced-constrained settings.

Several studies focused on individual differences in OSN privacy, e.g., the effect of gender~\cite{hoy2010} and age~\cite{Sheehan2002} on user behaviors. Kwon et al.~\cite{Kwon2019self_disclosure_OSN} studied the OSN users' self-disclosure activities and observed that they are more likely to disclose personal data when they can utilize positional advantages by playing bridging roles from their networks. Such et al.~\cite{such2017} studied particular MPCs over co-owned images from identification to resolution, and uncovered nuances and complexities, including co-ownership types, and divergences in the assessment of image audiences. Amon et al.~\cite{amon2020influencing} investigated the effects of several factors on decisions to share images of people on OSNs and found that developing interventions for reducing image sharing and protecting the privacy of others is a multi-variate problem.

The above previous privacy-related studies were dedicated to measuring or analyzing privacy issues, privacy settings, and factors that affect users' privacy-related behaviors. However, we did not see any past research conducting large-scale measurements of potential and actual privacy issues related to image sharing on OSNs, which is another gap our work will fill by presenting the first large-scale study of the problem of face privacy on Twitter.

\section{Understanding Face Privacy in OSNs}
\label{sec:overview}

In this section, we first define bystanders in the context of sharing images on OSNs. Then we discuss face privacy issues and relevant user behaviors in this context. Finally, we describe our methodology and introduce the datasets used in our study.

\subsection{Definitions: Subjects and Bystanders}
\label{subsec:bystander_defination}

To better understand the face privacy problem in the context of sharing images on OSNs from different perspectives~(i.e., bystanders and subjects), we randomly sampled 1,050 images shared on Twitter. Three authors of this paper jointly examined these images and found that 343 images~(32.67\%) contain at least one human face. We discussed potential privacy issues of those images and reached the agreement that for 277 images (26.38\%) there exist different types of potential privacy concerns, including unintended leakage of private information including faces, geo-location information, and other attributes of human subjects. We also found that, out of the 277 images, 232 images (83.75\%) had potential face leakage related to bystanders and/or subjects who are not the uploader, indicating that face privacy is among the most common privacy issues of such OSN images and that detecting bystanders and subjects in such images will help study different types of face privacy issues affecting different people.

To provide clear definitions of bystanders and subjects in OSN images, the three authors inspected 232 images with potentially leaked faces. We found that bystanders and subjects are complex concepts and their precise meanings are heavily context-dependent. For example, a bystander could be a person who was unaware of the image shooting or a person the photographer did not intend to capture. In some images, a person in the foreground occupies a large area and appears to be the photographer's intended target, but is not facing the camera, making it difficult to tell if the person was aware of the shooting activity. In~\cite{hasan2020bystander_privacy}, Hasan et al.\ define bystander as a person who is not a subject of the photo and is thus not important for the meaning of the photo, e.g., the person was captured in a photo only because they were in the field of view and was not intentionally captured by the photographer. Darling et al.~\cite{Darling2019,DarlingLL20} used a similar definition for bystanders: people who are captured inadvertently in others' pictures. The above definitions are framed from the photographer's perspective and do not take into account the implications for privacy protection.

Based on our analysis, we introduce the following definitions. A subject is \textit{a person who actively participated in an image-shooting activity}, and a bystander is \textit{a person who did not actively participate in an image-shooting activity}. By emphasizing active participation, our definition takes people's consent into account. Subjects actively engaged in image-capturing activities are likely aware of their inclusion in the photograph and may have consented to it, whereas bystanders may not be aware of or have given consent for their image to be used. This aligns with common privacy expectations, as people who willingly participate in image-shooting activities can reasonably expect their image to be captured and used, while bystanders in the background typically do not anticipate being included in photographs. Our definitions can effectively cover most of bystander cases we inspected and offer several advantages from a privacy protection standpoint as we will demonstrate in the following sections. In addition, visual cues in the photo can be used as good proxies for high-level concepts, such as willingness to be in the photo and actively posing for the photo, which has been confirmed in the work of Hasan et al.~\cite{hasan2020bystander_privacy}, giving us confidence that using visual cues in photos can represent \textit{active participation}. Since we now have a clear and precise boundary between subjects and bystanders and given that the features characterizing this definition can be extracted from the image, we can focus on the development of techniques to identify potential privacy concerns for humans involved in images.

\subsection{Face Privacy Issues of Bystanders and Subjects}
\label{subsec:objects}

To determine whether there is a face privacy issue with an OSN image, we ask the following two questions: (i) Did each person in the image agree to be photographed? (ii) Did they give their consent to the uploader to post the image on the OSN platform? Following our definitions of subjects and bystanders, we can argue that bystander(s) in an OSN image often did not realize that they had been photographed or did not give consent to the photographer, and it is more often that the photographer asked all subjects but not each bystander for their consent. Therefore, it is more likely that a bystander's privacy is violated compared to a subject's privacy, indicating the necessity and importance of focusing on bystander privacy.

We consider subjects to be active participants in image-shooting activities. While it is reasonable to assume they agreed to be photographed, it is not necessarily the case that they gave their consent for the image to be uploaded to an OSN platform. Subjects can include the uploader or friends (e.g., relatives, friends, colleagues, etc.) of the uploader on the OSN platform. If the uploader also appears in an OSN image uploaded, we can assume that there is no privacy issue about the uploader's privacy; however, for all other subjects, there is a possibility they may not like the image published therefore leading to a privacy concern. If a subject has a public profile with their face image, it does not mean that they are happy for their face appearing in all OSN images. For example, they may want to hide their social relationships with other people. Note that in this work we focus on ordinary people and exclude celebrities.

\subsection{Uploaders' Behaviors about Face Privacy}
\label{subsec:behav}

Before uploading an image to an OSN platform, the uploader can choose to anonymize some faces in the image to address privacy concerns. Previous studies have shown that using techniques like blurring to reduce recognizability in photos can increase individuals' willingness to be photographed and reduce privacy risks~\cite{dimiccoli2018mitigating, alharbi2019mask,DarlingLL20,KandappuSX21}. We classify such behaviors into the following three categories, with consideration for the detectability of faces in the photo by existing face detection models.
1) \textbf{No anonymization}: the uploader did not anonymize any faces in the image. In this case, without knowing if any of the non-uploader subject or bystander has a privacy concern, there is a \textit{potential risk of face privacy leakage}.
2) \textbf{Partial anonymization}: The uploader attempted some level of manipulation of one or more faces in the image so that \textit{some but not all} personally identifiable features on the faces are lost. If the photographed persons have privacy concerns, then there will be a \textit{certain risk of partial face privacy leakage} from the non-anonymized information, which may allow partial or even full re-identification of the affected individuals.
3) \textbf{Full anonymization}: the uploader manipulated the whole face so that it is impossible to re-identify the corresponding individual. In this case, there is \textit{no any face privacy risk}. Figure~\ref{figure_example-anonymized-photos}\footnote{All images with human faces in this paper were obtained from Unsplash, and our processing and use of images strictly follow the regulations of the website license: \url{https://unsplash.com/license}} shows an image with three categories of anonymization applied.

\newlength\sfigwidth
\setlength{\sfigwidth}{0.25\textwidth}
\begin{figure*}[tb]
\centering
\subcaptionbox{No anonymization}{\includegraphics[width=\sfigwidth]{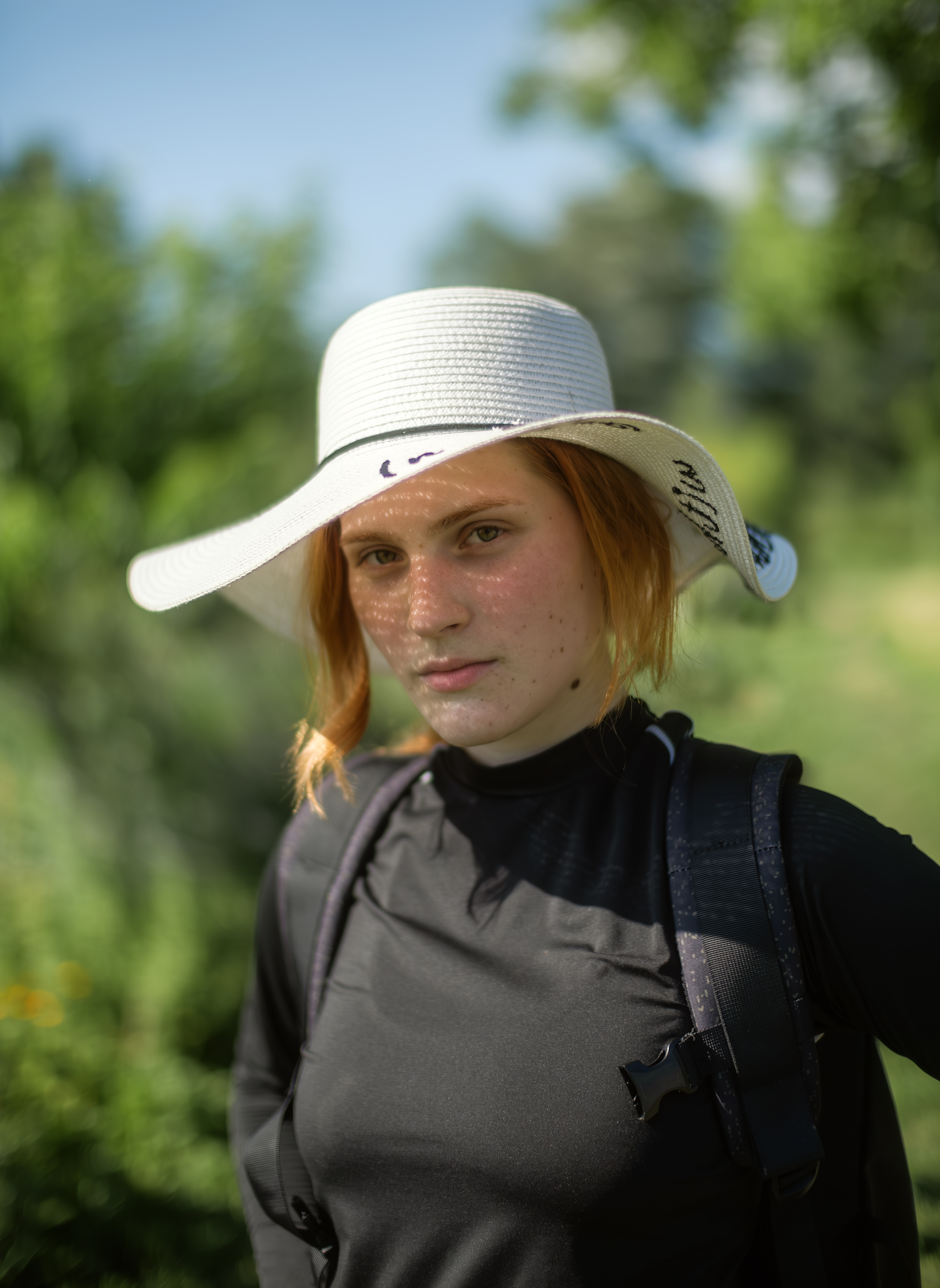}}
\hspace{1em}
\subcaptionbox{Partial anonymization}{\includegraphics[width=\sfigwidth]{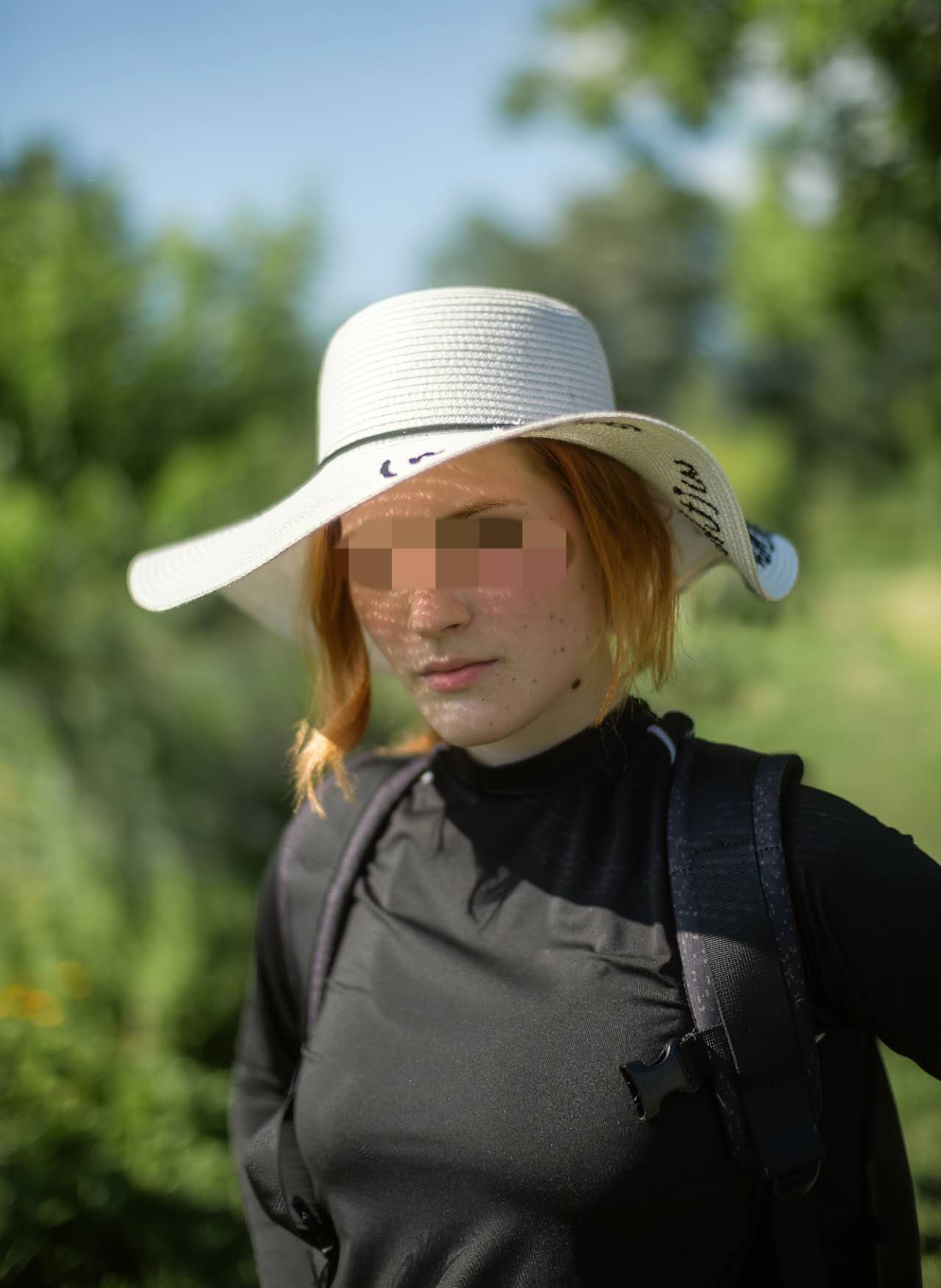}}
\hspace{1em}
\subcaptionbox{Full anonymization}{\includegraphics[width=\sfigwidth]{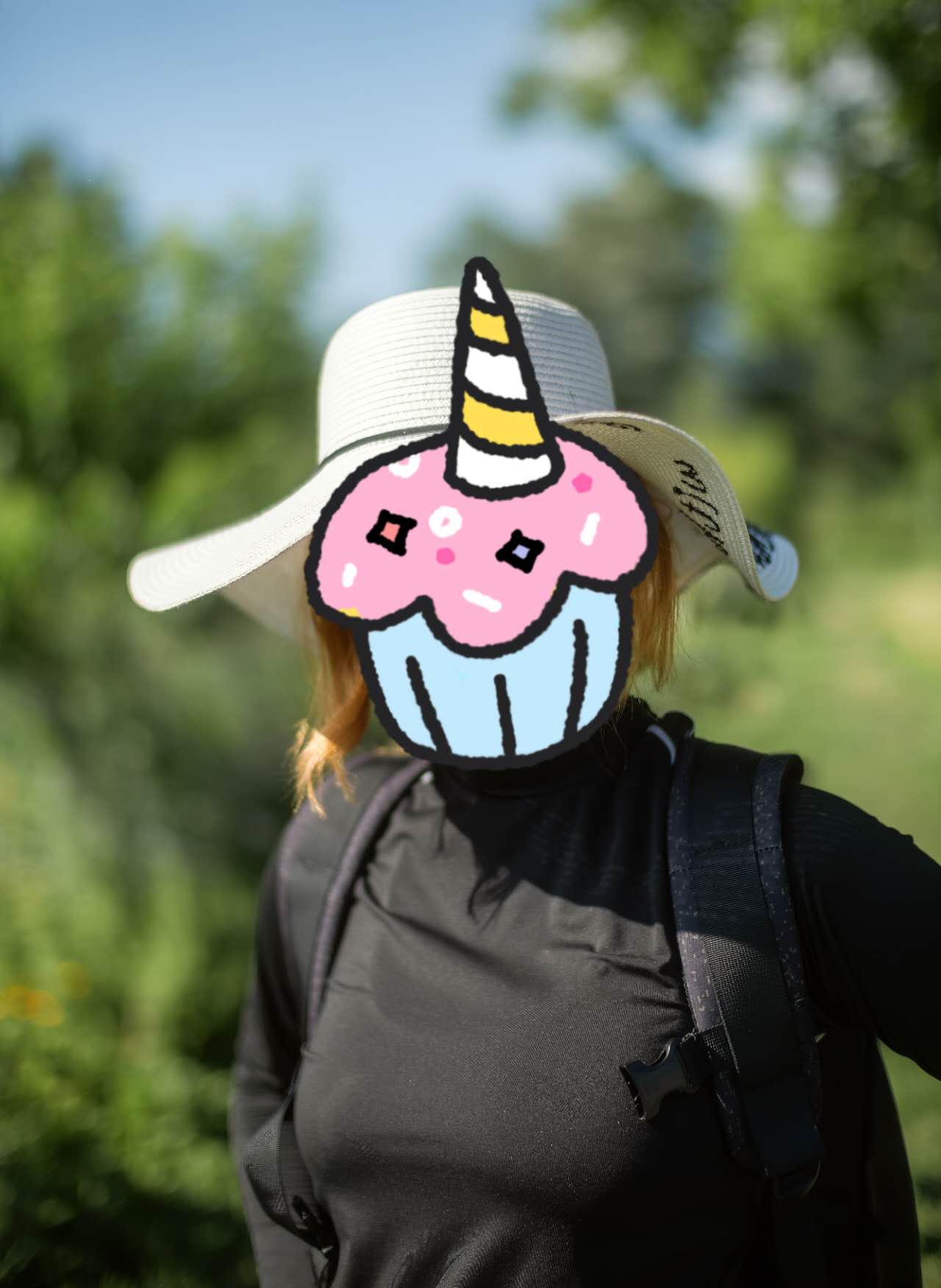}}
\caption{Examples of user anonymized,
partially anonymized, and fully anonymized faces.}
\label{figure_example-anonymized-photos}
\end{figure*}

\subsection{Advancing the State-of-the-Art}
\label{subsec:overall_Methodology}

As mentioned in Section~\ref{subsec:bystander_detection}, Hasan et al.~\cite{hasan2020bystander_privacy} and Darling et al.~\cite{Darling2019, DarlingLL20} are the only researchers who investigated the development of a bystander-subject classifier, but their work has the following three issues: 1) they did not apply their methods to large-scale and real-world OSN images; 2) they did not define bystanders clearly to take their privacy into account; and 3) the features they used do not fully characterize bystanders and subjects. Specifically, Hasan et al.'s algorithm works in the presence of the whole human body of each person in an image, while the face region features used by Darling et al.\ lack the association and contrast between individual face features and the features of the entire photo.

We address the first issue by constructing new datasets of OSN images and testing various methods on such images. The second issue is addressed via our new definitions of subjects and bystanders (Section~\ref{subsec:objects}) and uploaders' behaviors related to face privacy (Section~\ref{subsec:behav}). The third issue was evidenced by our inspection of 232 Twitter images with potential privacy issues: 201 images (86.63\%) contain at least one person whose whole human body was not captured in the image. Considering that faces are more consistently present in the Twitter images we inspected, we decided to develop our new classifier mainly based on features that can be derived from faces as we explain in Section~\ref{sec:sub_and_by}, similar to Darling et al.'s approach~\cite{Darling2019, DarlingLL20}. By using face-only features and carefully adding some other features, such as adding features like the comparison of face size to photo size, the comparison of face region blur to overall photo blur, and the number of people in the photo, our new bystander-subject classifier can address the third issue.

\subsection{Determining Subjects}
\label{subsec:Determining Subjects}

Since the uploader plays a special role in the analysis of face privacy, we adopted a heuristic rule to find out which subject is the uploader (Section~\ref{sec:pipeline}). Specifically, we compared the faces of all subjects in a target image with all faces appearing in the uploader's profile image. We did not consider the privacy of the faces that appeared in the users' profiles. This approach is grounded in the assumption that if a user utilizes their real face as their profile picture, there is no inherent privacy conflict when they post an image of themselves. In such cases, the uploader is the same as the face in the profile image, and the act of sharing their face is not a breach of privacy. On the other hand, if a user employs someone else's face as their profile picture, any potential privacy breach has already occurred at the point when they initially chose to use someone else's face as their avatar. Therefore, in our study, we temporarily treat the face in the profile as if it were the uploader's face. This approach is rational because it allows us to focus on privacy concerns related to the act of uploading a social media image, regardless of whether the profile picture corresponds to the user's real face. To understand the uploaders' face privacy protection behaviors, we used manual qualitative encoding to detect if the uploader manipulated any faces in each image we included in our large-scale analysis. The two types of manipulated faces (subjects or bystanders) and the degree of anonymization (no, partial, and full) are two aspects that we focused on during the encoding process and in our analysis.

\subsection{Face Privacy Datasets}

For our work, we constructed and used four new datasets as shown in Table~\ref{table_The_overview_of_our_datasets}. We describe the construction details of three datasets (Datasets 1, 2.A, and 2.B) in Section~\ref{subsubsec:Our_New_Subject-Bystander_Datasets} and the fourth one (Dataset 3) in Section~\ref{sec:privacy}.

\begin{table*}
\centering
\caption{The overview of our datasets}
\label{table_The_overview_of_our_datasets}
\begin{tabularx}{\linewidth}{lp{0.25\linewidth}cX}
\toprule
& Data Source(s) & \#(images) & Purpose\\
\midrule
Dataset 1 & public datasets, image sharing websites, Baidu, Douban, and Sina Weibo & 7,524 & Supporting the development (training, validation, testing and comparing with baseline models) of the bystander-subject classifier (Section~\ref{subsubsec:Experiment 1})\\
Dataset 2.A & Twitter & 496 & Testing the bystander-subject classifier's performance on OSN images (Section~\ref{subsubsec:Experiment 2} and Section~\ref{subsubsec:compare})\\
Dataset 2.B & COCO2017 dataset & 450 & Testing the bystander-subject classifier's performance on non-OSN images (Section~\ref{subsubsec:compare})\\
Dataset 3 & Twitter & 27,800 & Supporting the large-scale face privacy analysis on an OSN platform  (Section~\ref{sec:privacy})\\
\bottomrule
\end{tabularx}
\end{table*}

We chose Twitter as the only OSN platform for Datasets 2.A and 3 for the following three reasons. First, it is common for OSN-related research~\cite{Unintended_URLsndss/KaleliKENS21, uss/WeiSVRGHFWMU20, ndss/DrakonakisIIP19} to use Twitter as the only platform due to its large user base and/or open API at the time of data collection. Second, most other platforms with large user bases (e.g., Facebook) did not provide an open API when we conducted our work, therefore, they were less chosen by researchers for large-scale social media analytics work. Third, while some other OSN platforms such as Facebook may be less professional-facing and data-richer, due to the lack of an open API studying the problem on such platforms will require completely different data collection/analysis methods and consideration of more complicated ethical aspects, which will be left as our future work. 

\section{Subjects and Bystanders Classification}
\label{sec:sub_and_by}

This section gives details of our new bystander-subject classifier and explains how we evaluated its performance.

\subsection{Methodology}

We noticed that different attributes exist between face regions of bystanders and subjects: (i) subjects' faces are often larger, (ii) subjects are often in a more central position of the image, (iii) the photographer tends to focus on the subjects, and (iv) the subjects tend to face the camera. We use these representative attributes as features, including face size, face position, head pose, blurriness, and contrast. Besides, inspired by Hasan et al.~\cite{hasan2020bystander_privacy}'s work, we also added the face count as a feature. However, we did not use the gaze vector feature proposed by Darling et al.~\cite{Darling2019, DarlingLL20} because it is difficult to extract when the photographed person wore sunglasses. Our model works by following three steps: 1) face detection, 2) feature extraction, and 3) binary classification. These steps are described in detail in the rest of this section.

\subsubsection{Face Detection}
\label{subsubsec:face_detection}

We used RetinaFace~\cite{deng2019retinaface} to locate faces in an image. The coordinates of the face frame and eyes were recorded. All the features we extracted were limited within the face frame. The coordinates of the eyes were used to determine the face position of our feature extraction module.

\subsubsection{Feature Extraction}
\label{subsubsec:feature_extraction}

Intuitively, the promising factors that can be obtained from the image to classify subjects and bystanders include face size, face position, face count, pose, blurriness, and contrast. We extracted these factors with currently available techniques.

\textbf{Face size:} The subject is the target person and usually has a larger face size. We calculated the size of each face based on the coordinates of the face frame.

\begin{figure}[!t]
\centering
\includegraphics[width=0.6\linewidth]{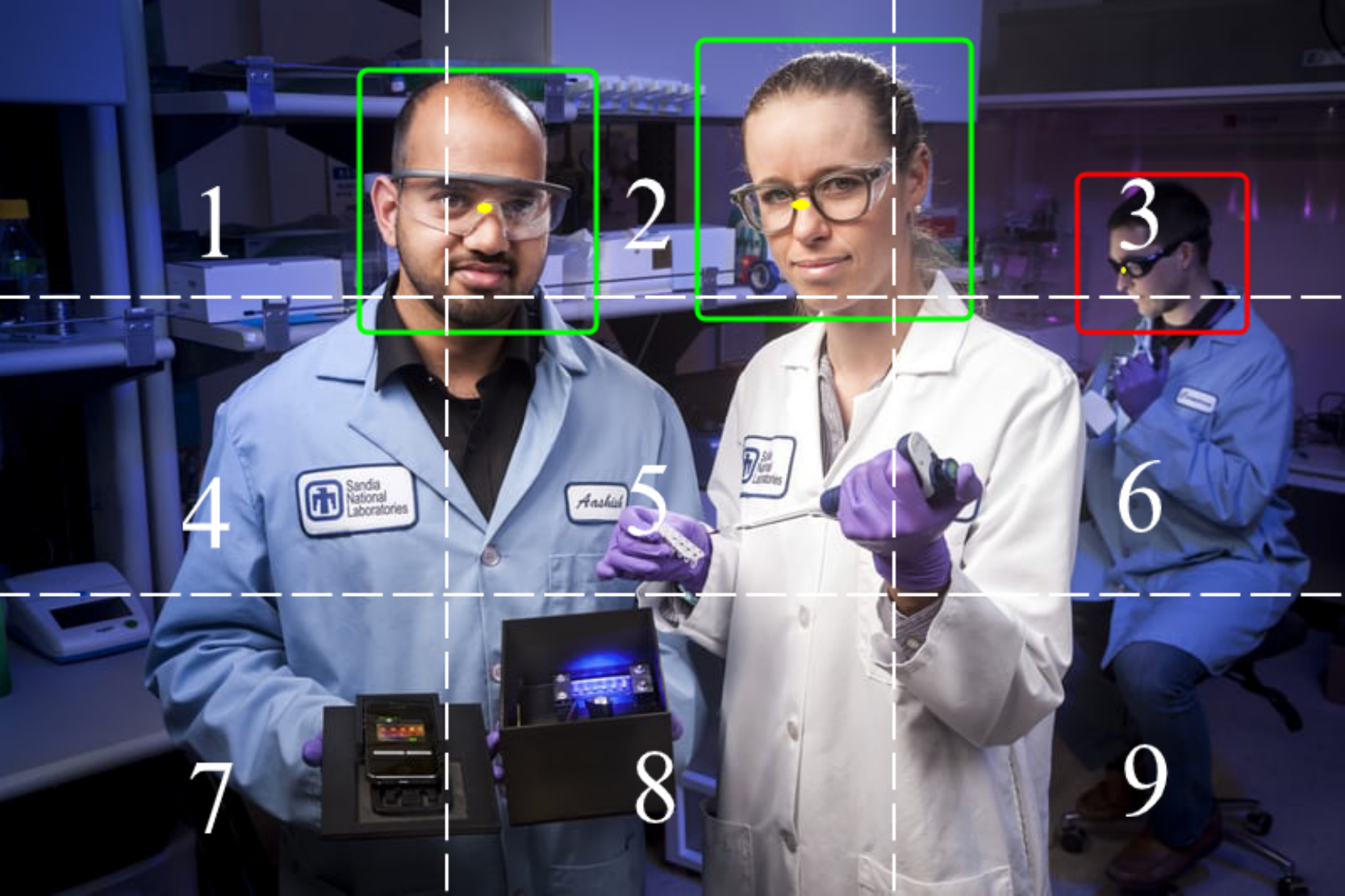}
\caption{An example image showing how faces in the images are positioned in one of the 9 regions and how to calculate the number of faces in each region. The two subjects (highlighted in the green box), who are clearly posing for the camera, are located in Region 2. The bystander (highlighted in a red box), who shows no indication of willingness to participate in the filming based on visual cues, is located in Region 3. Regions 1, 4, 5, 6, 7, 8, and 9 each have a face count of 0, while Region 2 has a face count of 2, and Region 3 has a face count of 1.}
\label{figure:example_face_position}
\end{figure}

\textbf{Face position and face count:} The subject is generally located in the center of the image. We divide the image into nine\footnote{We compared the influence of extracting location features and the number of people within each region on the classifier's performance. To do this, we experimented with image division into various configurations, including 4, 6, 9, and 16 regions. Our experiments revealed that the difference in accuracy between dividing the image into 9 and 16 regions was negligible. However, both of these configurations outperformed the results obtained when dividing the image into 4 or 6 regions. Therefore, we chose to divide the image into 9 regions.} equal-sized regions, numbered 1-9, as shown in Figure~\ref{figure:example_face_position}. We then calculated the midpoint between the eyes of each face, and the region containing this midpoint was used to represent the face's position as a feature. Additionally, we recorded the total number of faces in each image and the count of faces within each of the nine regions of the image. Note that a single face may cover multiple regions; in such cases, we use the region containing the eyes' midpoint to calculate the face count for each region.

\textbf{Head pose:} In most cases, the subject's pose for taking pictures is uniform, while bystanders may have different postures because they are not aware of the shooting. We used the pitch, yaw, and roll angle predicted by the head pose estimation algorithm proposed by Ruiz et al.~\cite{ruiz2018fine} to represent the pose of the face.

\textbf{Blurriness:} We used the value of the Laplacian operator to represent the blurriness of a given area. When the image is blurred, it contains less boundary information, and the variance of the corresponding Laplacian operator is small.

\textbf{Contrast:} It reflects the resolution of an image, and can be calculated according to Eq.~\eqref{equ:c}:
\begin{equation}\label{equ:c}
\text{Contrast} = \sum_{\delta}\delta(i,j)^2 P_{\delta (i,j)},
\end{equation}
where $\delta(i,j)=|i-j|$ represents the gray-scale difference between adjacent pixels, and $P_\delta(i,j)$ represents the probability of the gray-scale difference between adjacent pixels. The subject area usually has a higher resolution.

We extracted the above features for each face region. However, due to differences in photographers' preferences, equipment, shooting environment, etc., some raw data are not suitable to derive final features. For example, the face of a bystander in one image may be larger than the subject in another image. Therefore, we used the ratio of the face size to the image and to the largest face as features. Blurriness and contrast are also processed in this way. In addition, we used ResNet34~\cite{he2016deep} to extract image features of the face area. Specifically, we removed the last fully connected layer and took the feature map as image features. Table~\ref{table_feature_compare} shows the features used by Darling et al.~\cite{Darling2019, DarlingLL20} and Hasan et al.~\cite{hasan2020bystander_privacy} and the ones used in our work.

\begin{table*}
\centering
\caption{The comparison of features used by our bystander-subject classifier with those used by Darling et al.'s~\cite{Darling2019, DarlingLL20} and Hasan et al.'s~\cite{hasan2020bystander_privacy}}
\label{table_feature_compare}
\begin{tabularx}{\linewidth}{lX}
\toprule
Classifier & Features used\\
\midrule
Darling et al.'s~\cite{Darling2019, DarlingLL20} & size\_face / size\_face\_max, deviation of face from image's center, blurriness\_face, yaw, pitch, roll, gaze deviation\\
Hasan et al.'s~\cite{hasan2020bystander_privacy}
& size\_body / size\_image, predicted pose, replaceable, and photographer's intention (the last three derived from proxy features)\\
Ours
& size\_face / size\_image, size\_face / size\_face\_max, position\_face (one of 9 areas of the image), number\_of\_faces, number\_of\_faces in each of the 9 areas, yaw, pitch, roll, blurriness\_face / blurriness\_image, blurriness\_face / blurriness\_face\_max, contrast\_face / contrast\_image, contrast\_face / contrast\_face\_max, feature map extracted by ResNet-34\\
\bottomrule
\end{tabularx}
\end{table*}

We concatenated the feature map (512-dimensional vector) of the face region with face size, position, number, blurriness, contrast, and head pose (20-dimensional vector) as the final features, and used them as an input to our classifier.

\subsubsection{Binary Classification}

\begin{figure*}[!htb]
\centering
\includegraphics[width=\linewidth]{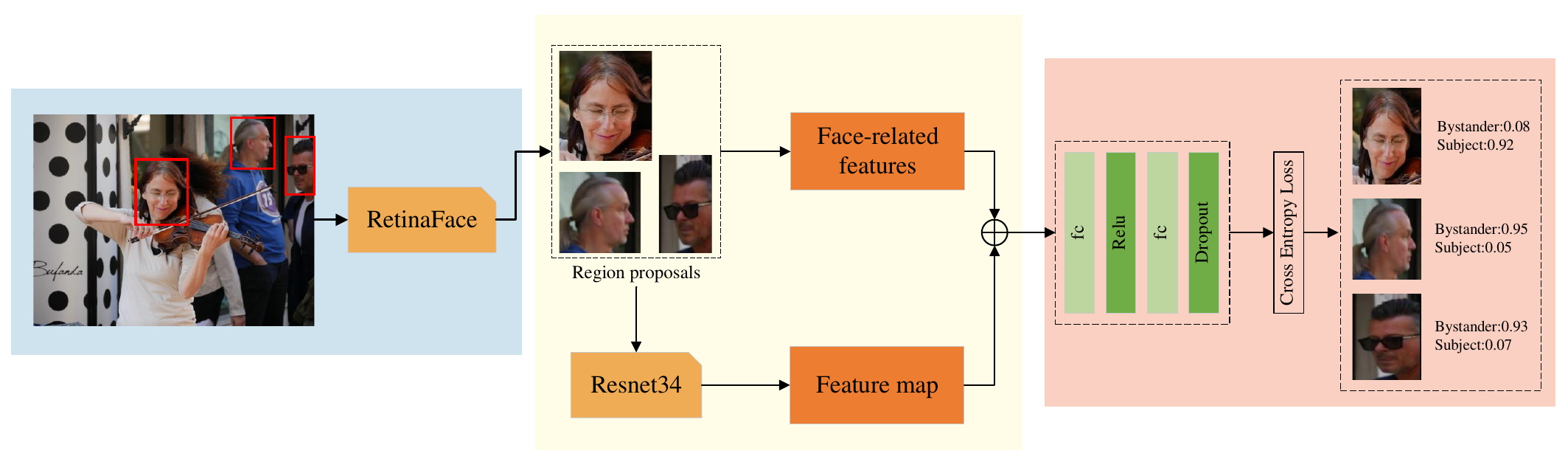}
\caption{An overview of our proposed bystander-subject classifier.}
\label{figure:overview-model}
\end{figure*}

Our classifier consists of two fully connected layers. The input of the first layer is the fused 532-dimensional feature. We used ReLU as the nonlinear activation function. To prevent overfitting and improve the generalizability, we added a dropout layer to reduce the number of parameters. The input of the second fully connected layer is a 128-dimensional feature vector, and the output is the binary classification result. We used Logsoftmax to convert the results into probability values. Figure~\ref{figure:overview-model} depicts an overview of our proposed bystander-subject classifier.

\subsection{Performance Evaluation}

We now explain the performance evaluation of our proposed bystander-subject classifier. Section~\ref{subsubsec:Our_New_Subject-Bystander_Datasets} describes three new bystander-subject datasets we constructed for this study: a larger dataset (Dataset 1) based on multiple data sources, and two small datasets -- Dataset 2.A solely based on Twitter to represent images on a typical OSN platform; and Dataset 2.B sampled from a non-OSN public image dataset. The following subsections explain three experiments that we have conducted. In Section~\ref{subsubsec:Experiment 1}, we compare our classifier with several baseline classifiers, all trained, validated and tested using Dataset 1. In Section~\ref{subsubsec:Experiment 2}, by using Dataset 2.A, we evaluate the performance of our classifier trained using Dataset 1. In Section~\ref{subsubsec:compare}, we compare our classifier with Darling et al.'s~\cite{Darling2019,DarlingLL20} and Hasan et al.'s~\cite{hasan2020bystander_privacy} classifiers using both Datasets 2.A and 2.B. Since Darling et al.\ demonstrated in their 2020 study~\cite{DarlingLL20} that the feature-based classifier they proposed in 2019~\cite{Darling2019} outperforms the CNN classifier, we have chosen to compare our method with their feature-based approach in this section. Considering that this is a binary classification task, we evaluate the performance of classifiers based on common metrics such as accuracy, precision, recall, F1-measure, TPR (True Positive Rate), and FPR (False Positive Rate). In our evaluation, we consider the bystander as the positive class and the subject as the negative one.

\subsubsection{Our New Subject-Bystander Datasets}
\label{subsubsec:Our_New_Subject-Bystander_Datasets}

\begin{figure*}[!htb]
\centering
\includegraphics[width=\linewidth]{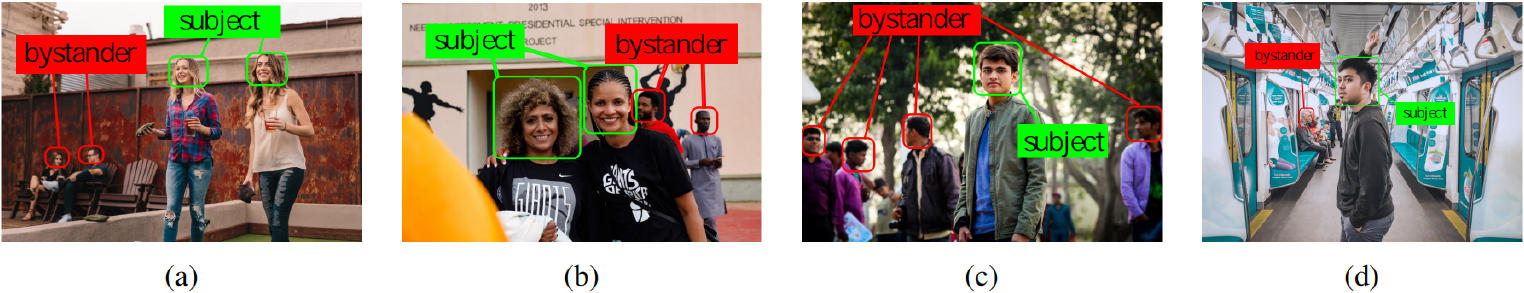}
\caption{Some example images in Dataset 1.}
\label{figure:example_images_Dataset1}
\end{figure*}

Due to the fact that the only two publicly available subject-bystander datasets are small and not diverse enough, we decided to construct a new general dataset (Dataset 1) for this study. To collect a diverse dataset of images containing people, we first manually selected images from three large publicly available face datasets (WIDER FACE~\cite{yang2016wider}, LFW~\cite{LFWTech}, and FDDB~\cite{fddbTech}). Then we used multiple web sources, including a web search engine (Baidu), four stock photography websites with a free license for research purposes (Unsplash~\cite{Unsplash}, Pexels~\cite{Pexels}, Pickupimage~\cite{Pickimage}, and Pixabay~\cite{Pixabay}), Douban~\cite{Douban}, a Chinese website with a large number of screenshots of film and television drama, and Sina Weibo~\cite{SinaWeibo}, a large OSN platform in China. In total, we collected 7,524 images containing faces, which cover rich shooting scenes such as restaurants, hospitals, streets, scenic spots, gyms, schools, companies, etc., and shooting activities, such as travel, interviews, elections, parades, dinners, sports, performances, etc. Figure~\ref{figure:example_images_Dataset1} shows some example images in Dataset 1, which consists of 22,369 subjects and 21,579 bystanders. Next, we used RetinaFace~\cite{deng2019retinaface} to locate face regions. After this step, we obtained 43,948 faces for annotations. 

\textbf{Annotation:} Due to the large amount of data to be annotated, we enlisted a third-party annotation company to complete the annotation task. We provided the company with 50 photos (containing 73 subjects and 63 bystanders) annotated by the first author of this paper as examples. Along with these examples, we gave clear guidelines for the annotators to examine the entire image and determine whether the person in each face frame was \textit{actively participating} in the photo. Annotators were asked to consider factors such as the context of the scenario, the individual's pose, and their position within the photo. Cases that were ambiguous were marked for further review by the first author of the paper. When 10\% of the labeling task was completed (i.e., 760 photos had been labeled), we randomly selected 70 photos and had them independently labeled by the first author. We calculated Cohen's kappa~\cite{landis1977measurement} between the company and the first author. They reached an inter-annotator agreement of 0.83, indicating the annotators understood the task and followed the guidelines. Following this assessment, the annotation company proceeded to complete the remaining labeling tasks. After all images in Dataset 1 were annotated by a third-party company, the first author randomly selected 1,000 images from the dataset and annotated them independently. We also used Cohen's kappa to measure the inter-rater reliability between the annotation results from the third-party company and those of the first author, resulting in a score of 0.72, indicating a fair degree of consistency.

In addition, we sampled 496 real-world images containing unanonymized human faces from Twitter to construct Dataset 2.A. Due to the dataset's size, three co-authors of this paper conducted the labeling process. People in photos were labeled as bystanders only when two or all of the three co-authors agreed. We extracted a total of 4,156 faces, including 1,567 subjects and 2,589 bystanders. The consensus rate among the three co-authors annotating this dataset, measured using Fleiss' kappa, is 0.79.

Finally, we also randomly sampled 450 non-OSN images containing human faces from the COCO2017 dataset~\cite{Lin2014cocodataset} to construct Dataset 2.B. The same three authors labeled images for Dataset 2.A did the labeling work for the 450 images and people in photos were labeled as bystanders only when both or all three authors agreed. We extracted a total of 1,748 faces, including 866 subjects and 882 bystanders. The Fleiss' kappa score of this dataset is 0.82.

\subsubsection{Experiment 1}
\label{subsubsec:Experiment 1}

To verify that the features we used can best discriminate between subjects and bystanders, we designed three baseline models that also focus on the face area to compare with our proposed scheme. Formally, the compared methods are as follows: 1) \textbf{MaskRCNN~\cite{he2017mask}:} It is a popular object detection network. We trained this model to directly classify the faces of subjects and bystanders. 2) \textbf{Feature map (FM):} We only used ResNet-34~\cite{he2016deep} to extract the image features of the face region and classify subjects and bystanders. 3) \textbf{Face-related features only (FF):} We used only face size, number of faces, face position, blurriness, contrast, and head pose as features. 4) \textbf{All features (FM+FF):}\footnote{We conducted a comparative analysis of classification results using LR, SVM, and XGBoost. The utilization of a two-layer neural network demonstrated higher classification accuracy and F1 scores in comparison.} We used the face size, number of faces, face position, blurriness, contrast, and head pose as well as the feature map as final features.

\begin{figure*}[!htb]
\centering
\subcaptionbox{Correct classifications}{\includegraphics[width=\sfigwidth]{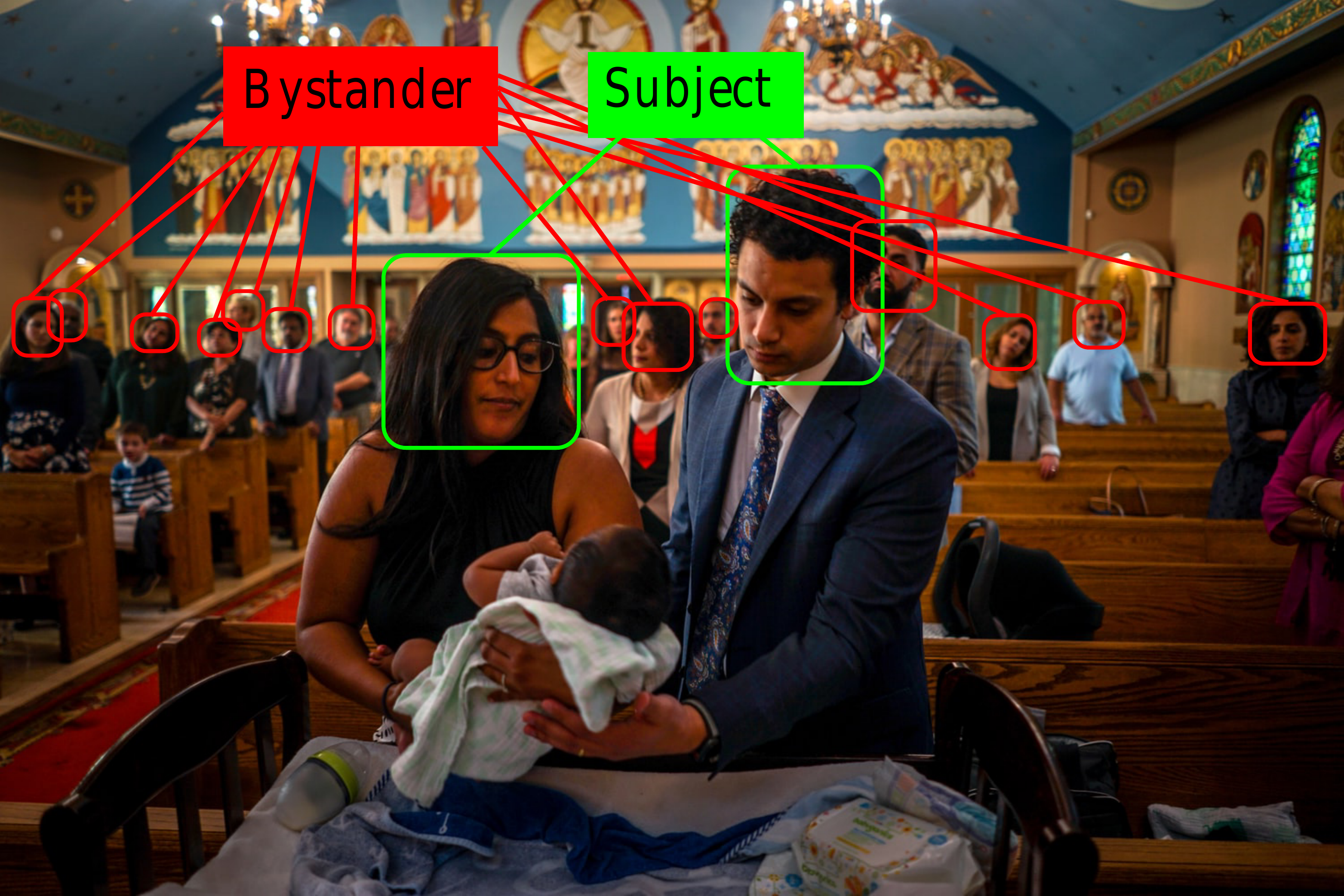}}
\hspace{1em}
\subcaptionbox{2 false-positive errors}{\includegraphics[width=\sfigwidth]{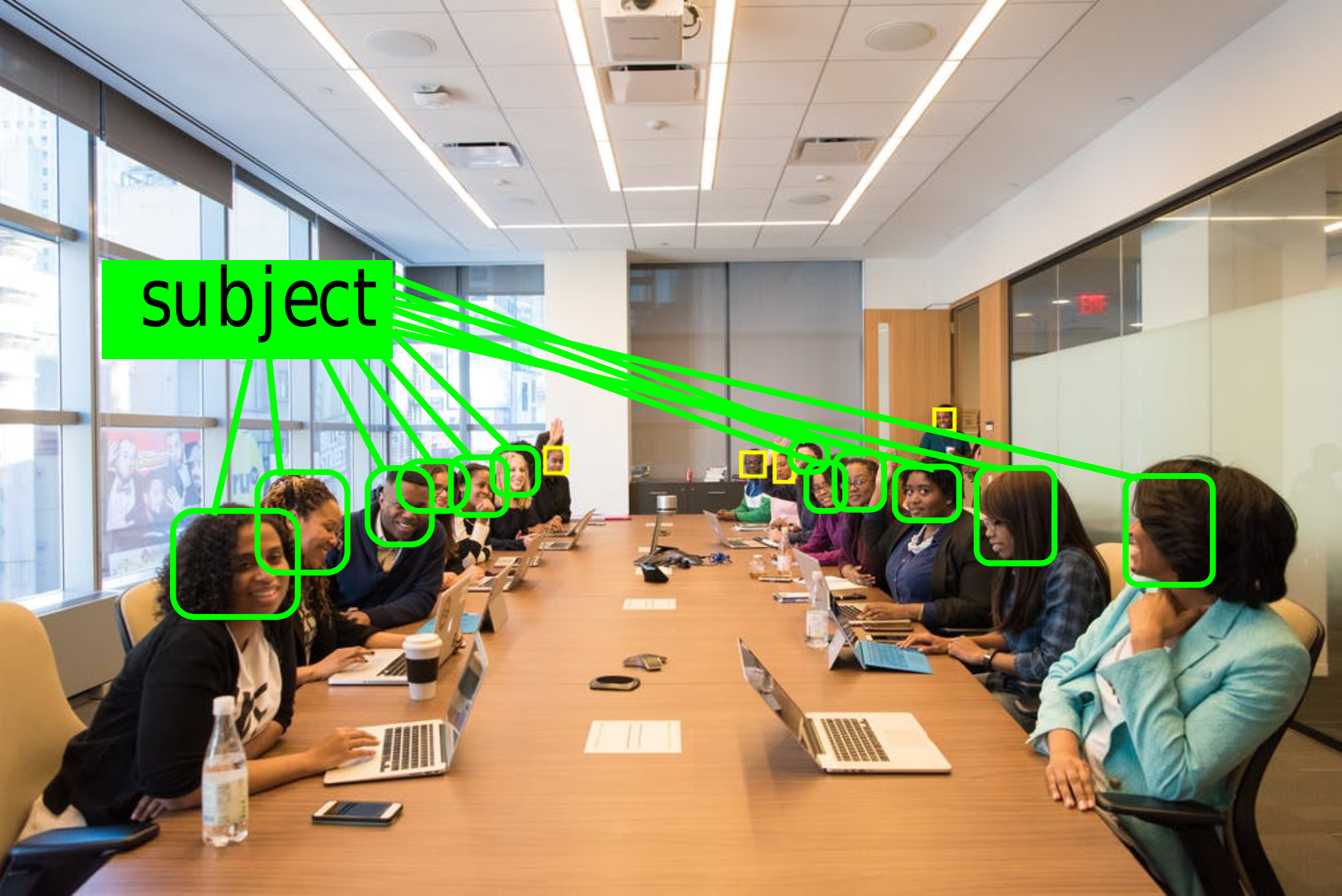}}
\hspace{1em}
\subcaptionbox{1 false-negative error}{\includegraphics[width=\sfigwidth]{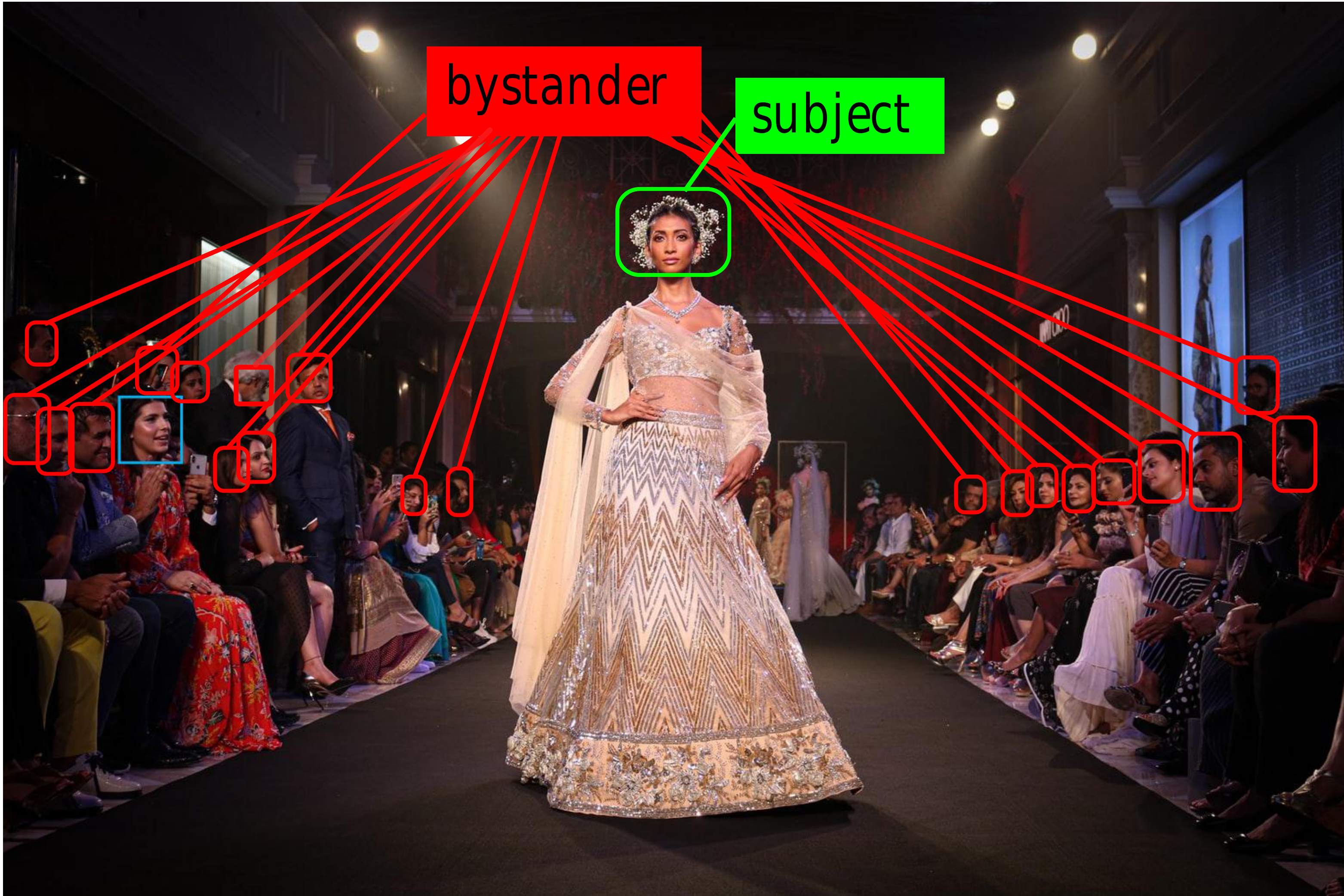}}
\caption{Examples of correct and incorrect classification results: red boxes -- correctly classified bystanders; green boxes -- correctly classified subjects; yellow boxes -- subjects that are misclassified as bystanders; and blue boxes -- bystanders that are misclassified as subjects.}
\label{figure_example-classified-photos}
\end{figure*}

We used Dataset 1 and performed an 80-10-10 split into the training, validation, and test sets. Table~\ref{table_Metric_scores} shows the results based on the test set. The performance of both MaskRCNN and the feature map is far worse than the face-related features and all features. This mirrors findings from Darling et al.~\cite{DarlingLL20}, who similarly observed higher accuracy with feature-based models compared to CNN models using direct face region inputs. Additionally, in our experiments, the accuracy and recall of face-related features are slightly lower than all features. These results indicate that face-related features are more indicative and important than image features of the face regions in the task of classifying subjects and bystanders. Figure~\ref{figure_example-classified-photos} shows the correct case, false positive case, and false negative case of our model on the test set, respectively.

\begin{table*}[!htb]
\centering
\caption{Metric scores of baselines and our proposed model in Dataset 1.}
\label{table_Metric_scores}
\begin{tabularx}{\linewidth}{l YYYYY}
\toprule
Method & Acc & R/TPR & P & F1 & FPR\\
\midrule
MaskRCNN~\cite{he2017mask} & 73.62\% & 63.71\% & 74.92\% & 68.86\% & 18.01\%\\
FM & 76.16\% & 66.41\% & 81.97\% & 73.37\% & 14.30\%\\
FF & 93.03\% & 91.08\% & \textbf{95.09\%} & 93.04\% & \textbf{4.94\%}\\
\textbf{FM+FF} & \textbf{94.48\%} & \textbf{93.57\%} & 94.07\% & \textbf{94.58\%} & 5.61\%\\
\bottomrule
\end{tabularx}
\end{table*}

We would like to further compare our work with that of~\cite{hasan2020bystander_privacy, Darling2019, DarlingLL20} on Dataset 1. However, the method proposed in~\cite{hasan2020bystander_privacy} needs to crop the whole person's body in the image. The number of people included in Dataset 1 is large, so manual cropping is difficult. Additionally, the accuracy of existing algorithms for detecting people in images is lower than that for detecting faces, so using tools for detecting people to automatically crop images cannot guarantee that the obtained people can correspond one-to-one with the faces in our Dataset 1. To address the above issues, we used Datasets 2.A and 2.B to simultaneously compare these schemes (Section~\ref{subsubsec:compare}).

\subsubsection{Experiment 2}
\label{subsubsec:Experiment 2}

We trained our model with all data in Dataset 1 to get the final classifier. To verify if our method could generalize, we further evaluated the classifier's performance based on Dataset 2.A. This experiment consists of two parts: 1) we took all the images as input to verify that this classifier can achieve reasonably high accuracy on OSNs images. 2) we divided images by the number of subjects in each image and obtained image groups with the number of subjects 1, 2, 3, 4, 5, 6-10, and above. Then we observed the performance when the number of subjects changed.

\textbf{Results and Analysis:} Our model achieves high scores in accuracy (95.00\%) and Recall/TPR (98.18\%), indicating that it has advantages in detecting bystanders. 47 (3.00\%) of the 1,567 subjects and 156 (6.03\%) of the 2,589 bystanders are classified in error. We checked these images and identified two situations: 1) one subject in the image is too prominent (for example, the face is too large relative to other people) so that the features of other subjects become similar to bystanders; and 2) one image contains a large number of subjects.

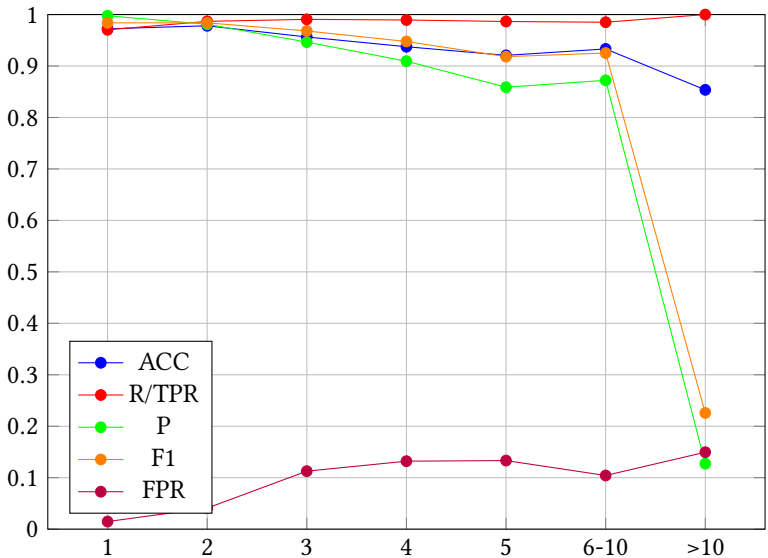
\begin{figure}[!htb]
\centering
\begin{tikzpicture}
    \begin{axis}[
        width=0.8\textwidth,
        height=0.6\textwidth,
        xlabel={},
        ylabel={},
        ymin=0, ymax=1,
        ytick={0,0.1,...,1},
        xtick={1,2,3,4,5,6,7},
        xticklabels={1,2,3,4,5,6-10,>10},
        legend pos=south west,
        grid=major,
    ]
    
    \addplot[color=blue,mark=*] coordinates {
        (1,0.9723) (2,0.9782) (3,0.9564) (4,0.9375) (5,0.9207) (6,0.9332) (7,0.8537)
    };
    \addlegendentry{ACC}

    \addplot[color=red,mark=*] coordinates {
        (1,0.9704) (2,0.9868) (3,0.9907) (4,0.9894) (5,0.9865) (6,0.9851) (7,1)
    };
    \addlegendentry{R/TPR}

    \addplot[color=green,mark=*] coordinates {
        (1,0.9977) (2,0.9811) (3,0.9465) (4,0.9094) (5,0.8588) (6,0.8722) (7,0.1273)
    };
    \addlegendentry{P}

    \addplot[color=orange,mark=*] coordinates {
        (1,0.9839) (2,0.9839) (3,0.9681) (4,0.9477) (5,0.9182) (6,0.9252) (7,0.2258)
    };
    \addlegendentry{F1}

    \addplot[color=purple,mark=*] coordinates {
        (1,0.0148) (2,0.0409) (3,0.1127) (4,0.1321) (5,0.1333) (6,0.1043) (7,0.1495)
    };
    \addlegendentry{FPR}
    
    \end{axis}
\end{tikzpicture}
\caption{Our proposed bystander-subject classifier's performance w.r.t.\ the number of subjects.}
\label{figure_Model-performance}
\end{figure}

Figure~\ref{figure_Model-performance} shows the results of the second experiment. As the number of subjects in an image increases, accuracy and Recall/TPR still maintain high scores, indicating that our model has a stable ability to detect bystanders. The changing trends of precision and F1 scores are the same. 
In addition, when the number of subjects is less than 11, the score of each indicator is better than or close to the average score, indicating that our model has advantages when the number of subjects is small. The performance degradation of this model is mainly due to misclassifying subjects as bystanders. The reason is that our Dataset 1 lacks images with a large number of subjects.

In the previous subsection, we observed that the accuracy using only face-related features is similar to all features. Therefore, we also examined the variation of the accuracy of only face-related features with the number of subjects. As shown in Table~\ref{table_Accuracy_two_features}, when the number of subjects is small, the accuracy rates of the two schemes are close. But when the number of subjects exceeds 10, using only face-related features will cause a significant drop in accuracy.

\begin{table*}[!htb]
\centering
\caption{Accuracy of our model on images with different number of subjects.}
\label{table_Accuracy_two_features}
\begin{tabular}{lccccccc}
\toprule
Number of subjects & 1 & 2 & 3 & 4 & 5 & 6-10 & $>$10\\
\midrule
FF & 95.60\% & 95.30\% & 91.74\% & 88.91\% & 85.37\% & 88.31\% & 80.18\%\\
\textbf{FF+FM} & \textbf{97.22\%} & \textbf{97.80\%} & \textbf{95.64\%} & \textbf{93.75\%} & \textbf{92.07\%} & \textbf{93.32\%} & \textbf{85.37\%}\\
\bottomrule
\end{tabular}
\end{table*}

\subsubsection{Experiment 3}
\label{subsubsec:compare}

We compared our method with the model proposed by Darling et al.~\cite{Darling2019, DarlingLL20} and Hasan et al.~\cite{hasan2020bystander_privacy} based on both Datasets 2.A (OSN images) and 2.B (non-OSN images). Neither Darling et al.\ nor Hasan et al.\ have released their source code or pre-trained models. Therefore, we reproduced Darling et al.'s face feature-based classifier as described in~\cite{Darling2019,DarlingLL20}. We also reproduced Hasan et al.'s classifier based on their description in~\cite{hasan2020bystander_privacy} and the public dataset they released. Since Hasan et al.'s features are whole body-based, we used YOLO~\cite{huang2018yolo} to detect human bodies in the input image, and then we manually added 32 human bodies with visible faces that YOLO failed to detect. We performed 10-fold cross-validation for both classifiers and on both datasets. The results are shown in Table~\ref{table_Accuracy_10-fold}.

\begin{table*}[!htb]
\centering
\caption{Comparison of performance of our model vs. Darling et al.~\cite{Darling2019,DarlingLL20} and Hasan et al.~\cite{hasan2020bystander_privacy} on 10-fold cross validation}
\label{table_Accuracy_10-fold}
\begin{tabular}{lcccccc}
\toprule
\multirow{2}{*}{} & \multicolumn{3}{c}{Dataset 2.A (OSN images)} & \multicolumn{3}{c}{Dataset 2.B (non-OSN images)}\\
\cline{2-7} 
& \multicolumn{1}{c}{\multirow{1.3}{*}{\centering Darling et al.'s}} & \multicolumn{1}{c}{\multirow{1.3}{*}{\centering Hasan et al.'s}} & \multicolumn{1}{c}{\multirow{1.3}{*}{\centering Our}} & \multicolumn{1}{c}{\multirow{1.3}{*}{\centering Darling et al.'s}} & \multicolumn{1}{c}{\multirow{1.3}{*}{\centering Hasan et al.'s}} & \multicolumn{1}{c}{\multirow{1.3}{*}{\centering Our}}\\
\midrule
ACC & 92.7\% & 82.7\% & \textbf{95.8\%} & 87.3\% & 75.3\% & \textbf{93.2\%}\\
P & 91.4\% & 88.1\% & \textbf{97.3\%} & 88.3\% & 70.6\% & \textbf{94.3\%}\\
R/TPR & 92.0\% & 84.7\% & \textbf{95.8\%} & 86.3\% & 78.5\% & \textbf{92.5\%}\\
F1 & 91.7\% & 86.5\% & \textbf{96.2\%} & 87.2\% & 74.2\% & \textbf{92.3\%}\\
FPR & 13.1\% & 21.0\% & \textbf{4.4\%} & 12.2\% & 27.2\% & \textbf{6.0\%}\\
\bottomrule
\end{tabular}
\end{table*}

\textbf{Analysis:} Our classifier achieved an average accuracy of 95.8\% on Dataset 2.A (OSN images), while Hasan et al.'s method is only 82.7\%, representing an improvement of 13.1\%. In addition, our classifier achieved an average accuracy of 93.2\% on Dataset 2.B (non-OSN images), while Hasan et al.'s method is only 75.3\%, representing an improvement of 17.9\%. The results showed that our classifier works well with both OSN and non-OSN images. There are at least two possible reasons why Hasan et al.'s classifier did not work as well as ours. First, the features it uses are less ideal. As discussed in Section~\ref{subsec:overall_Methodology}, many images involving privacy issues, especially those on OSNs, have partially occluded human bodies due to many reasons, so sometimes it can be difficult or impossible to extract the whole human body. Second, the amount of data used to train their three models for predicting features is small and the data are not well-aligned with data on OSN platforms. Therefore, when the trained models are directly applied to images on social media or image sharing websites, the three predicted features could be less accurate, which can then further affect the final performance of the classifier. Third, to solve the above-mentioned second problem, more training data will have to be collected and the classifier retrained. This will require recruiting human annotators who will have to look at a set of collected images and provide four labels for each bystander in each image in the new training set: bystander determination, pose evaluation, replaceability assessment, and the photographer's intention. Note that the latter two labels are quite subjective so the human annotators will have to spend more time to consider. In contrast, if the platform wants to further improve the (already better) performance of our classifier, they just need to recruit human annotators to indicate one binary label for each bystander: if it is a bystander or a subject. Obviously, the human annotators' efforts involved for our classifier are much simpler (1 vs 4) and more objective (just 1 binary label indicating if a face belongs to a bystander), therefore can be done much faster and with less subjective bias. We expect that the human labeling effort for Hasan et al.'s classifier is at least four times more expensive than that for our classifier.

Compared with Darling et al.'s method, our classifier improves accuracy by 3.1\% on dataset 2.A (OSN images) and 5.9\% on dataset 2.B (non-OSN images). Although both our classifier and Darling et al.'s are based on face features, their features have certain limitations. We found that the gaze deviation feature could not be extracted in cases where the face was too small or the subject was wearing sunglasses. In addition, they did not capture contrast features between the subject and the overall photo, such as the proportion of the face in the photo, the number of people in the photo, etc. Despite these limitations, the accuracy of both our method and Darling et al.'s, which are based on facial features, is higher than that of Hasan et al.'s method, which is based on whole-body features.

\section{Semi-automated Framework for Analysis of the Face Privacy Problem}
\label{sec:pipeline}

\begin{figure*}[!htb]
\centering  
\includegraphics[width=\linewidth]{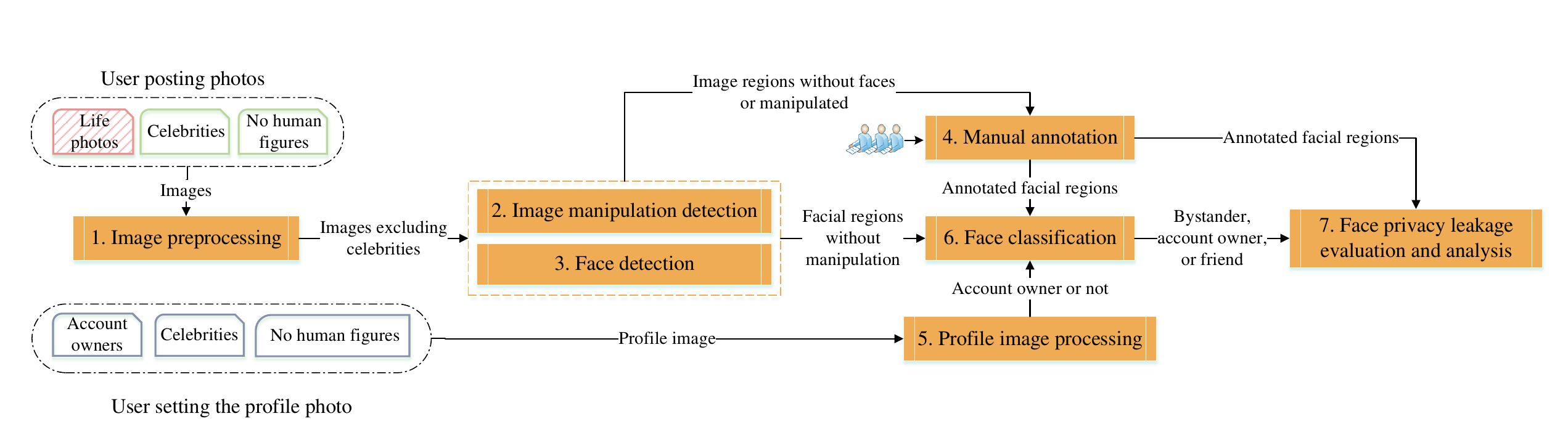}
\caption{Overview of our proposed framework.}
\label{figure_Pipeline}
\end{figure*}

Based on the classification model, we propose a semi-automated framework for quantitative and qualitative analysis of the face privacy problem on social media platforms. Figure~\ref{figure_Pipeline} depicts our proposed framework. To illustrate the framework's effectiveness and to provide meaningful insights, we selected Twitter as the example OSN platform, which is a mainstream microblogging platform that allows users to upload an image as their profile image and post tweets with images. It is essential to underscore that our framework exhibits adaptability and can be applied to other OSN platforms. In the following, we describe each step of the framework in detail.

\textbf{Step~1 -- Image preprocessing:} Facial images uploaded online may contain celebrities such as actors, politicians, and athletes. Such images are often widely circulated on online platforms for the benefit of the celebrities who are usually happy to see such publicity and do not have privacy concerns. Therefore, for our proposed framework we decided to exclude celebrities and focus on normal people. Our framework first employs the Google image reverse search tool~\cite{GoogleImages} to exclude such images and then provides an interface for human users of the framework to screen the remaining images to eliminate those featuring only celebrities. Google Reverse Image Search is instrumental in locating the source of an image, thereby enabling the verification and retrieval of the associated contextual information. This approach helps in identifying images that feature celebrities and have been disseminated online. To mitigate the possibility of false positives inherent in reverse image searches, a rigorous validation process was implemented. The first author of this article conducted a meticulous manual review of all the identified images. During this review, images in which every photographed individual was confirmed to be a celebrity were excluded from the analysis.

\textbf{Steps~2 and 3 -- Image manipulation detection and face detection:} Uploaders may have modified the face region in an image before uploading it. Such manipulation behaviors directly reflect awareness and actual action taken by the uploaders, so they are very important to be covered by our framework. For Step~2, our framework utilizes MVSS-Net proposed by Dong et al.~\cite{dong2021mvss} to detect possible manipulations applied to images from Step~1. MVSS-Net can automatically detect and localize pixel- and image-level manipulations (e.g., copy-move, splicing, and inpainting). In parallel (Step~3), our framework uses a face detection algorithm to locate faces in the image. The whole image can be categorized into four types of regions: 1) facial regions without manipulation; 2) facial regions with manipulation; 3) manipulated regions without a face; and 4) regions without face or manipulation.

\textbf{Step~4 -- Manual annotations:} Various factors such as light and human posture can cause some faces to be undetected and their corresponding regions labeled as Type 4 (no face or manipulation). To capture such missed faces, manual inspection of images is required. Additionally, for regions labeled as Type 3 (manipulated but without a face) manual inspection is required to determine whether the manipulation is for a person. For modified face regions (Type 2, 3, and 4), manual inspection can be used to determine which parts have been modified, how they have been modified, and the potential modification intentions of the uploader. To facilitate such manual annotations of images, we developed an encoding scheme as part of our framework, based on the work of three co-authors of this paper who tried to encode many images in the large-scale analysis we will report in the next section. The codes cover the following aspects: face verification, face manipulation verification, manipulation intention\footnote{See Appendix~\ref{appendix:Manipulation Intention} for more information and examples of how we inferred the potential manipulation intention.}, facial part manipulated, and manipulation method (Table~\ref{table_coding_scheme}). 
An encoding protocol was agreed to ensure the quality of the encoding results. The three annotators first determined whether each Type 3 region contains a face, and for each confirmed facial region manipulation-related codes were determined. One of them initially coded 50\% of Types 2, 3, and 4 regions, after which the three annotators refined the coding scheme and reached an agreement on all coded regions in a group discussion. Then, they coded all image regions independently. All coding disagreements were resolved by consensus in a group discussion. For Types 3 and 4 regions that they agreed that there is a face, one of them annotated the face region using Labelme~\cite{russell2008labelme}. The annotators removed images that do not contain any faces since they provide no useful information on face privacy. Finally, all faces were classified into three classes: \textbf{Class A:}  un-manipulated faces, \textbf{Class B:} partially manipulated faces that remain detectable by the automated facial recognition tool used in our framework, and \textbf{Class C:} fully manipulated faces that were not detected by the automated facial recognition tool.

\textbf{Step~5 -- Profile image processing:} 
As discussed in Sections~\ref{subsec:objects} and \ref{subsec:Determining Subjects}, it is important for our framework to capture faces in profile images. To this end, our framework uses RetinaFace~\cite{deng2019retinaface} to detect if the profile image contains one or more faces. For profile images containing faces, our framework compares facial features of the face with those of celebrity faces in GIPHY's open-source Celebrity Detection Deep Learning Model~\cite{GIPHY2020celeb-detection-oss}. Meanwhile, we used Twitter API~\cite{TwitterAPI:search} to check if the account being checked is verified. With all the checks, our framework considers non-celebrity faces of non-verified uploaders as those who use their own faces as the profile image.

\textbf{Step~6 -- Face classification:} This step classifies each face in an OSN image into three categories: the uploader (i.e., the account owner), other subjects other than the uploader (e.g., the uploader's friends\footnote{In the rest of the paper, for the sake of simplicity, we will use the term \textbf{friends} for all such subjects. The term ``friends'' represents \textit{individuals who are non-uploaders but actively engage in the image-shooting process}. The use of the term ``friends'' in this context does not pertain to social link friends, as encountered on OSN platforms.}), and bystanders who are not the uploader. 
The framework first replaces each Class C face with a marked face to incorporate it in the later pipeline, and then our proposed classifier in the previous section is used to classify all faces in each image into subjects and bystanders. When the uploader uses their real facial image as their profile image (as identified in Step~5), our framework compares each face with the one in the uploader's profile image to further classify it into one of the three above-mentioned categories. To facilitate further discussions in the rest of the paper, we use \textbf{bystander*} to refer to \textit{a bystander in an image who is not the uploader}.

\textbf{Step~7 -- Face privacy leakage evaluation and analysis:} After Step~6, further analysis of faces in all collected images can be done to produce various statistics and to gain useful insights about the face privacy problem on the target OSN platform. These analyses can be based on the three classes of faces: 1) \textbf{Class 1} faces that are not the uploader themselves, which are not manipulated/anonymized so may leak privacy; 2) \textbf{Class 2} faces are insufficiently manipulated for privacy or other purposes -- if any of the three facial parts, eyes, the nose, and the mouth, are manipulated, we consider the face partially anonymized since manipulating such key facial parts can make it harder to recognize the face, therefore, may have some effect of privacy protection
; and 3) \textbf{Class 3} faces are fully manipulated/anonymized, which helps to protect face privacy.

\begin{table*}[!htb]
\centering
\caption{The coding scheme for image regions without faces and those with manipulated faces}
\label{table_coding_scheme}
\begin{tabularx}{\linewidth}{llX}
\toprule
& Code & Description\\
\midrule
Face Verification & Contain faces & There are recognizable faces in the image area or it can be inferred from the image context that the area originally contains faces.\\
& No faces & There are no discernible faces in the image area and no one can be inferred from the image context.\\
\midrule
Manipulation Verification & Face manipulation & Manipulations to the face region affect identity recognition.\\
& No face manipulation & The face is not modified or does not affect identification.\\
\midrule
Manipulation Intention
& Privacy & Prevent faces from being identified.\\
& Humor & Entertain viewers or expressing irony.\\
& Beauty & Hide facial imperfections or spice up images.\\
& Information & Convey information to viewers, such as mood, identity information, etc.\\
& Unknown & The intention cannot be inferred based on image context.\\
\midrule
Manipulation Part(s) & Whole body & Faces, clothing, and body movements.\\
& Whole face & Face but not clothing or body movements.\\
& Eye & Eyes but not the whole face.\\
& Nose & Nose but not the whole face.\\
& Mouth & Mouth but not the whole face.\\
& Ear & Ears but not the whole face.\\
& Others & Cheeks, forehead, etc., but not the whole face.\\
\midrule
Manipulation Method & Blur & Softening the selected region by blur obfuscation.\\
& Pixel & Applying pixel obfuscation to the selected region.\\
& Mask & Masking the selected regions with stickers, cartoon figures, other faces, or color blocks.\\
& Distort & Wrapping the selected regions.\\ 
\bottomrule
\end{tabularx}
\end{table*}

\section{Face Privacy Analysis of Twitter Images}
\label{sec:privacy}

In this section, we use our proposed framework described in the previous section to perform a large-scale study on 303,801 OSN images from Twitter. As highlighted in Section~\ref{sec:pipeline}, our framework is designed to operate across a range of social media platforms, extending beyond Twitter. We first describe our data collection process in Section~\ref{subsec:Real-world_Twitter_Data_Collection}. Subsequently, in Section~\ref{subsec:Data_Results_of_Uploader_Posted_Faces}, we report the data results of uploader posted face images, non-anonymized faces, and anonymized faces. Then we report our findings~\ref{subsec:Findings} from four aspects, involving general uploader behaviors and facts about face privacy on Twitter (Section~\ref{subsec:General_Behaviors_Facts_Face_Images}), behaviors of uploaders who did not anonymize faces (Section~\ref{subsec:Behaviors_Users_not-protect_faces}) and of those who chose to do so (Section~\ref{subsec:Behaviors_Users_protected_faces}), and new evidence about potential leakage of social attributes from images containing leaked faces (Section~\ref{subsec:Leakage_Social_Attributes}).

\subsection{Real-world Twitter Data Collection}
\label{subsec:Real-world_Twitter_Data_Collection}

Our methodology requires two types of data: images posted by users (i.e., uploaders) and the uploaders' profile images. 
We utilized the Twitter Streaming API~\cite{TwitterAPI:streaming} to sample tweets at three different time points -- 7:00, 15:00, and 23:00 -- on both a typical working day (Friday, July 2, 2021) and a non-working day in most countries (Saturday, July 3, 2021). This approach allowed us to capture data from diverse users in various countries and regions, as we collected data at different times on both working and non-working days. We sampled tweets for five minutes on the working day and for one minute on the non-working day, resulting in a total of 3,344 and 3,230 uploaders' tweets, respectively. After removing duplicate user IDs, we retained data from 6,569 users. Then, we used the Twitter API~\cite{TwitterAPI:search} to collect uploaders' profile images and the latest 50 images posted, taking into account the inherent constraints of the Twitter API when it comes to accessing user historical data. If an uploader posted fewer than 50 images, we included all images from their timeline. 146 users did not post any images, and we excluded these users from our analysis. We collected images from the remaining 6,423 users. Among these users, 514 posted fewer than 50 images, and we collected all images they had posted. The remaining 5,909 users each uploaded at least 50 images. For these users, we collected their 50 latest images. In total, we collected 303,801 images. We sampled users speaking over 20 languages, with English-speaking users (51.41\%), Japanese-speaking users (13.62\%), and Spanish-speaking users (9.13\%) being the three largest user groups.

\subsection{Data Results of Uploader-Posted Faces}
\label{subsec:Data_Results_of_Uploader_Posted_Faces}

\subsubsection{Results of Subjects and Bystanders}
\label{subsubsec:Results_of_Subjects_and_Bystanders}

Our proposed framework can automatically detect and classify faces in images based on the predefined strategy. Among the 6,423 uploaders who posted images, 78.78\% of posted images contain at least one face. These images can be categorized into images containing celebrities (e.g., advertisements, posters, news reports, movie stills, and screenshots) and real-world images of non-celebrity people. 3,860 uploaders posted at least one image containing one or more celebrity faces and 3,036 ones posted at least one image containing one or more non-celebrity faces. As previously mentioned, for our work we focused on face privacy in the latter category as celebrity-related images are rarely considered privacy concerns. After filtering out the former type and images without any faces through Steps~1--5, we found that 46.22\% of uploaders (3,036/6,569) posted a total of 27,800 real-world images containing at least one non-celebrity face, and 57.21\% of these uploaders (3,036) published at least five images containing one or more non-celebrity faces. Our framework further classified these real-world faces (Step~6).

We report classification results at three levels: face, image, and uploader. \textbf{Face level}: We detected 83,782 faces from the 27,800 images (i.e., our \textbf{Dataset 3}), including 53,942 subjects and 29,840 bystanders. The data results are shown in Table~\ref{table:number_of_faces_for_subjects_bystanders} (Appendix~\ref{appendix:tables}). Among them, friends and bystander*\footnote{We defined ``friends'' and ``bystander*'' in Step~5 of our framework described in Section~\ref{sec:pipeline}. } are at potential risk of face privacy leakage. We used the face similarity detection tool reported in~\cite{Deng_2019_CVPR} to compare all faces posted by uploaders to count unique faces.\footnote{Due to the large number of faces, it is more complicated to compare all the faces, so we adopted a compromise method. We first randomly sampled 10\% of uploaders who have posted real-world facial images. We then compared the similarity of all faces posted by these uploaders and detected 543 unique faces, of which 7 (1.31\%) faces were repeated among different uploaders. We examined these 7 duplicate faces and concluded that they were false positives. This suggests that the same face is unlikely to appear in images posted by randomly sampled uploaders. Therefore, we counted the unique faces posted by the same uploader and simply summed up the number of unique faces of all uploaders as the final number of unique faces. In addition, we considered each Class 3 face published by an uploader as a unique face because the original face is unavailable.} We detected 38,081 unique subjects (70.60\% of all detected friends) and 28,507 unique bystanders (95.53\% of all detected bystanders*). Our analysis shows that, unlike the frequent appearance of the same subject in the images shared by uploaders, bystanders rarely appear in different images of one uploader since they are mostly just random ``non-targeted'' strangers to the photographer. \textbf{Image level}: 74.10\% of the images contain only one or more subjects, 0.54\% contain only one or more bystanders, and the remaining 25.36\% contain both at least one subject and at least one bystander. This shows that most of the images focused on subjects, but some images still include bystanders who should not have been captured. We further classified all people into the uploader, the uploader's friends, and bystanders*. The images containing only the uploader (16.38\%) do not normally have face privacy issues, but the remaining (83.62\%) of the images may leak the privacy of other subjects or bystanders. \textbf{Uploader level}: Among the 3,036 uploaders who posted images containing faces, 31.62\% of them posted images contain only subjects, 0.03\% posted images contain only bystanders, and the remaining 68.35\% posted images contain both at least one subject and at least one bystander. We found that only 1.58\% of uploaders posted images containing only the uploader themselves, which poses no privacy risks. We report image- and uploader-level results in Table~\ref{table:Basic_statistics_of_user_posting_photos} (Appendix~\ref{appendix:tables}).

\subsubsection{Results of Anonymized Faces}

We report data results of uploader anonymization of faces at the same three levels as before: face, image, and uploader. \textbf{Face level:} Our pipeline (Step~7) showed that 89.23\% (74,758/83,782) of detected faces had not been anonymized in any way. As discussed in Section~\ref{sec:overview}, such faces can lead to potential privacy issues. Only 0.68\% (573/83,782) of detected faces were fully anonymized, and there is no risk of privacy leakage for these people. In addition, 0.12\% (99/83,782) of detected faces were partially anonymized and 70 of them were modified for privacy anonymization, and they suffered certain partial privacy breaches. Detailed data are provided in Table~\ref{table_Face-level_privacy_leakage/protection_statistics} (Appendix~\ref{appendix:tables}). \textbf{Image level:} 82.65\% (22,976/27,800) of images contain non-anonymized faces, 0.23\% (65/27,800) contain partially anonymized faces, and 1.20\% (338/27,800) contain fully anonymized faces. We classified the images that may have privacy issues based on the degree of anonymization and the category of the face, resulting in a total of 26 image categories, as shown in Table~\ref{table:photo-level_privacy_leakage+protection_statistics} (Appendix~\ref{appendix:tables}). Of these categories, 19 had potential face privacy leakage, 11 had certain partial privacy leakage, and 3 had no privacy leakage. \textbf{Uploader level:} 97.63\% (2,964/3,036) of uploaders posted images with non-anonymized faces, 1.42\% (43/3,036) posted images with at least one partially anonymized face, and 5.70\% (172/3,036) posted images containing at least one fully anonymized face. We classified the uploaders who posted images with privacy problems based on the degree of anonymization and the category of the face, resulting in a total of 24 uploader categories, as shown in Table~\ref{table:user-level_privacy_leakage+protection_statistics} (Appendix~\ref{appendix:tables}). Of these categories, 19 showed potential face privacy leakage behaviors, 13 had certainly partial privacy leakage behaviors, and 2 had no privacy leakage behaviors.

As mentioned before, our proposed framework can detect whether faces within an image have been partially or fully manipulated, which may indicate the uploader's intention of anonymizing the faces for privacy purposes. In total, we identified 234 uploaders who manipulated at least one facial part, a whole face, or a human body of 766 photographed people (including 652 subjects and 114 bystanders*) within 474 images. Out of these uploaders, 194 partially or fully anonymized 672 persons (566 subjects and 106 bystanders*) appear in 400 images. In Tables~\ref{table:protected_bystander_face} and \ref{table:protected_subject_face} (Appendix~\ref{appendix:tables}), we list our inferred intentions of the uploaders who manipulated faces. The inference was done based on the consensus understanding of each manipulated face among the three authors who did the annotation work, judged based on the context of the image posted. Among all manipulated faces, for 81.46\% the intention was judged to be about privacy protection. Among all fully and partially anonymized faces, 7.14\% of them were considered for other intentions, although these faces could be considered anonymized to some extent due to the manipulation of crucial facial features. Faces of bystanders* that were anonymized constitute only 0.36\% of all such faces. Anonymized faces of friends account for 1.24\% of all such faces. Only 1.74\% of images containing faces of at least one friend and 0.69\% of images containing at least one bystander* had at least one face anonymized. Only 1.84\% (38 out of 2,068) of uploaders who posted one or more images containing at least one bystander* face anonymized them, and only 6.36\% (189 out of 2,973) of uploaders who posted one or more images containing at least one friend's face anonymized them. Among these few uploaders who actively anonymized faces, we discovered three new findings that are certainly privacy-related and we report these three findings in Section~\ref{subsec:Behaviors_Users_protected_faces}.

\subsection{Findings}
\label{subsec:Findings}

\subsubsection{General Uploader Behaviors and Facts}
\label{subsec:General_Behaviors_Facts_Face_Images}

The results reported in Section~\ref{subsubsec:Results_of_Subjects_and_Bystanders} can reveal the following \textbf{basic behavioral characteristics of uploaders who posted images containing faces}. 1) Users shared a wide variety of face-related images on Twitter. 2) Many uploaders posted not only images containing their own faces but also those containing their friends and bystanders, which can pose potential privacy risks. 3) The number of times the subject is repeated in images posted by a single uploader (i.e., the same subject appears in multiple images of the same uploader) is higher than that of bystanders.

In addition to these general observations, we conducted a more detailed analysis to categorize the types of bystanders appearing in the images. \textbf{Finding 1: After examining images containing at least one bystander, we discovered multiple subcategories of face privacy scenarios that have never been reported before.}
We randomly sampled 200 images containing at least one bystander and identified the following four subcategories: 1) unaware bystanders who did not know about the image-taking (e.g., passers-by); 2) unwilling bystanders who expressed reluctance to be photographed (e.g., alcoholics); 3) dis-empowered bystanders who understood the circumstances that they may be photographed and images may be uploaded (e.g., shop and restaurant staff and owners); and 4) secondary uploaders who had certain access to the camera (e.g., family members of the photographer). As per our definition in Section~\ref{subsec:bystander_defination}, these are individuals who do not actively participate in the shooting, but the level of consent given by each category of bystander during the photo-shooting and image-uploading process varies. Recently Zheng et al.~\cite{Zheng2022non-consensual_photo} designed an automated classifier capable of detecting unconsciously photographed individuals (i.e., unaware people) in images. However, there are currently no techniques available to detect other categories of people. Moving forward, a more refined classifier should be developed to establish more specific privacy protection policies for each subcategory of bystanders.

\subsubsection{Behaviors of Uploaders Who Did Not Anonymize Faces}
\label{subsec:Behaviors_Users_not-protect_faces}

\textbf{Finding 2: Uploaders do not actively anonymize faces in images, especially when the faces belong to bystanders.}
Unanonymized bystanders* account for 99.64\% of the total number of bystanders*. Unanonymized friends account for 98.76\% of all friends. 98.75\% of images containing friends have non-anonymized friends and 99.23\% of images containing bystanders have non-anonymized bystanders. 99.22\% of uploaders who posted images of bystanders containing non-anonymized bystanders, and 99.09\% of uploaders who posted images of friends containing non-anonymized friends. At the uploader, image, and face level data, it is evident that most uploaders do not actively anonymize individuals with potential privacy risks, with the anonymization of bystanders being even lower than that of friends. This finding highlights the fact that uploaders are not safeguarding faces with potential risks, particularly bystanders. Although this finding may not be surprising, it is the first time concrete quantitative evidence is given based on real-world data analysis, to the best of our knowledge. The reasons for such behavior require further investigation through interviews, surveys, and other methods in the future.

\textbf{Finding 3: Account type and profile image type are related to uploaders' non-anonymization behavior.}
Among uploaders who published real-world facial images, 98.65\% of verified uploaders have no anonymization for friends, which is higher than that of ordinary accounts (96.82\%). Furthermore, 84.14\% of verified uploaders have no anonymization for bystanders, much higher than ordinary accounts (62.52\%). Since verified accounts are more influential on social media, images posted through them may be forwarded by their followers, leading to a larger and more diverse audience. Note that we are not suggesting that verified accounts leak more face privacy of bystanders because it is possible that the bystanders in these images are followers of celebrities who have given permission to post the images. However, the true intentions of these bystanders cannot be inferred from the images alone, and research methods such as survey and interview are necessary to draw conclusions. Moreover, we found that uploaders using a real face as their profile image have a lower percentage of not anonymizing friends (96.18\%) than the other two types of uploaders (98.47\% and 98.22\% for no human figures and celebrities, respectively). On the other hand, they have a higher percentage of not anonymizing bystanders (70.22\%) than the other two types of uploaders (61.12\% and 57.99\% for no human figures and celebrities, respectively). We also ran a number of chi-square tests to check if the differences are statistically significant and the results are shown in Table~\ref{table_Chi-square_test}, which confirm the statistical significance of all differences observed at the significance level of 0.05. This finding suggests that we could implement privacy alerts or enforce policies for certain types of accounts to better protect face privacy on OSNs.

In this subsection, we present the limited uploader behaviors associated with non-anonymized images. We want to emphasize that the relationship between these behaviors and privacy leaks is uncertain, and some of these behaviors may not be related to privacy. Our measurement methods may not provide specific ratios, but they help shed light on the overall trends of uploader behavior.

\begin{table}[!htb]
\centering
\caption{Results of $\chi^2$ tests.}
\label{table_Chi-square_test}
\begin{tabular}{ccccc}
\toprule
& \multicolumn{2}{c}{Account} & \multicolumn{2}{c}{Profile image} \\ 
& $\chi^2 $ & $p$ & $\chi^2$ & $p$\\
\midrule
Friend & 5.115 & 0.024 & 13.532 & 0.001\\
Bystander* & 89.571 & $<$0.001 & 30.488 & $<$0.001\\
\bottomrule
\end{tabular}
\end{table}

\subsubsection{Behaviors of Uploaders Who Chose to Manipulate Faces}
\label{subsec:Behaviors_Users_protected_faces}

\textbf{Finding 4: Inadequate anonymization of faces may lead to insufficient privacy protection.}
In all 624 faces that were manipulated for privacy intentions, our framework detected 70 faces from these regions due to insufficient anonymization, which cover 21 bystanders* and 49 friends. It indicates that some uploaders were aware of privacy issues related to faces but may not have had sufficient knowledge on how to implement sufficient protection, leaving the possibility of partial re-identification of the face. Tables~\ref{table_manipulation_parts} and \ref{table_manipulation_methods} show different facial parts and methods used by uploaders to manipulate subjects and bystanders*' faces, respectively. We found that insufficient anonymization of friends mainly stemmed from uploaders manipulating only some key facial parts, such as the eyes, the nose, and/or the mouth, but not the whole face. In contrast, the partial anonymization of bystanders* is mainly due to insufficient blurring of their faces. Furthermore, as shown in Table~\ref{table_manipulation_parts}, there are 78 face regions manipulated, but the ears were never anonymized, indicating that uploaders did not have the knowledge on ear biometrics that can allow full or partial re-identification of people~\cite{Ganapathi_2023_ear_recognition}.

\begin{table*}[!htb]
\centering
\caption{The numbers of faces for different manipulation parts (privacy intention).}
\label{table_manipulation_parts}
\begin{tabularx}{\linewidth}{l YYYY}
\toprule
& \multicolumn{2}{c}{Bystander} & \multicolumn{2}{c}{Subject}\\
\midrule
Part & Partial Anonymization & Anonymization & Partial Anonymization & Anonymization\\
Body & 7& 9 & 1 & 25\\
Face & 6 & 63 & 3 & 333\\
Eye & 8 & 2 & 26 & 31\\
Eye, nose, mouth, others & & 5 & 3 & 70\\
Eye, nose, mouth & & & 1 &\\
Eye, nose, ear, others & & & 1 &\\
Eye, nose, others & & & 1 & 5\\
Eye, nose & & & 1 & 2\\
Eye, others & & 1 &\\
Mouth, nose, others & & & 11 & 6\\
Mouth, others & & & 1 &\\
Nose & & & & 1\\
Nose, others & & & 1 &\\
\bottomrule
\end{tabularx}
\end{table*}

\begin{table*}[!htb]
\centering
\caption{The number of faces for different manipulation methods (privacy intention).}
\label{table_manipulation_methods}
\begin{tabularx}{\linewidth}{l YYYY}
\toprule
 & \multicolumn{2}{c}{Bystander} & \multicolumn{2}{c}{Subject}\\
 \midrule
Method & Partial Anonymization & Anonymization & Partial Anonymization & Anonymization\\
Blur & 13 & 19 & 4 & 48\\
Pixel & & 13 & 1 & 33\\
Mask & 8 & 47 & 44 & 391\\
Distort & & & &\\
Blur \& mask & & & & 3\\
\bottomrule
\end{tabularx}
\end{table*}

We also observed that uploaders tended to fully or partially anonymize faces rather than the entire body, even though the latter could also provide good information for partial or unique re-identification of the individual (e.g., tattoos, highly unique clothes, hairstyle, bag(s), pet(s), and the body shape), therefore leading to privacy issues beyond faces. To demonstrate this additional privacy risk, let us give some real-world examples. In our collected images, we selected nine unique faces posted by nine uploaders that were partially anonymized in some images but not anonymized in others. We analyzed the images with anonymized and non-anonymized faces of the same person and were able to re-identify the identity of the anonymized faces based on the scene and the clothes worn by the individual. This observation is not surprising given that anonymizing the whole human body is not a common practice, e.g., Google Street View~\cite{GoogleStreetView} only blurs faces. The privacy issues related to whole human bodies in images deserve further research.

\textbf{Finding 5: Uploaders who intend to preserve the privacy of others may also fail to protect all faces from privacy risks within images due to their inconsistent behaviors.}
In 363 images that were manipulated for privacy intentions (as judged by the three annotators), we found that 124 of them still contain fully leaked faces, indicating that the uploader did not manipulate all faces with privacy implications in each of such images. There are two different situations. In one situation (for 109 out of the 124 images), the uploader only focused on the privacy of a particular sub-group of photographed persons, such as anonymizing faces of friends while ignoring faces of bystanders* (for 106 images) or the other way round (for 3 images). In the other situation (for 100 out of the 124 images), the uploader anonymized some but not all of the same type of photographed persons. In the second situation, there may be different reasons for the lack of anonymization of all photographed persons of the same type: for friends, this may be because only some friends of the uploader requested their faces to be anonymized so the uploader decided to leave the others' faces untouched; for bystanders*, the fact that some faces were anonymized indicates that the uploader had the awareness/intention to protect bystanders*' privacy in general, although for various reasons they did not anonymize all (e.g., overlooked some, or simply felt it too time-consuming to anonymize all bystanders*). Furthermore, our framework detected only 45 out of 175 (25.7\%) privacy-conscious uploaders modified all the images posted, while others only anonymized faces in some images posted. Only 20 uploaders (11.4\%) fully anonymized the face of each person in each image. Such inconsistent behaviors can clearly lead to privacy issues for some photographed persons.

\textbf{Findings~4} and \textbf{5} together indicate that most uploaders failed to anonymize faces due to insufficient manipulation of privacy-sensitive regions in images and inconsistent manipulation behaviors. To address such issues, we recommend that automated uploader-facing tools, such as our bystander detection classifier, should be deployed on OSN platforms to help warn uploaders about more inadequate anonymization for all photographed persons before they upload an image.

\textbf{Finding 6: The use of third-party mask faces for manipulation purposes may result in further privacy breaches.}
When performing Step~4 of the framework, the three annotators observed six uploaders using a third-party real face as a mask to cover the original facial regions for nine images, of which two faces were manipulated for privacy purposes only, four for humor purposes only, two for the above two purposes, and one for unknown reasons. The mask faces used for two images with humor intentions are popular memes on the web, and the identity of the remaining seven faces could not be determined by the annotators. By using a mask face, the six uploaders anonymized the faces in the original images, but they may have compromised the privacy of the owners of the mask faces if consent was not obtained. This observation requires further investigation in future work, with a large number of cases and with an empirical study involving human participants to better understand how people see the use of mask faces for different purposes and what privacy concerns people could have. To address the potential privacy issues of using real mask faces, AI-generated ``deepfake'' faces can be used instead, although more research is also needed to ensure such AI-generated faces do not leak privacy of real faces in the training data of the AI-based face generator.

\subsubsection{Potential Leakage of Privacy-Sensitive Social Attributes}
\label{subsec:Leakage_Social_Attributes}

In the process of detecting image scenes and analyzing face privacy issues, we discovered potential leaks of privacy-sensitive social attributes of photographed persons.

\textbf{Finding 7: More personal sensitive information can be inferred from non-manipulated faces.}
We used the pre-trained scene recognition model reported in~\cite{Place365} to detect the scene of each image collected, which often reflects where the image was captured. From 23,020 images with potential or certain privacy issues, we identified a wide range of venues that can be potentially privacy-concerning, which include hospitals (740, 3.21\%), nursing homes (536, 2.33\%), army bases (154, 0.67\%), conference rooms (60, 0.26\%), churches (19, 0.08\%), youth hostels (16, 0.06\%), and drugstores (11, 0.05\%). Venues can reveal privacy-sensitive information about people, e.g., army bases and conference rooms are closely related to profession, churches can reveal the religious belief of a person,  youth hostels are related to people's age and economic status, and drugstores, hospitals, and nursing homes may reveal people's health conditions. It seems that uploaders did not normally consider such more subtle privacy leaks.

\textbf{Finding 8: Social relationships between the uploader and some photographed persons can be inferred.}
We analyzed the frequency of unique faces and their similarity to the uploader who used a real face as their profile image. Specifically, we examined the images posted by 1,621 uploaders and found two different types of information leakage that can help infer social relationships between the uploader and the photographed persons. First, for 1,538 uploaders some non-anonymized faces appeared more than once across multiple images. The higher frequency of such faces can often indicate that the photographed individuals have a close social relationship with the uploader, e.g., being a family member or having a romantic relationship. Examining the image timelines of all the uploaders, the three annotators were able to confidently infer the romantic relationship between the uploader and one photographed person from 24 images. Second, 613 uploaders posted images containing faces that are not identical to but moderately similar to the uploader's face. By examining the scene of each of such images, the three annotators were able to easily infer the parent-child relationship between the uploader and one of the photographed persons for 1,571 images of 70 uploaders. Note that such inferences were made from the image alone without examining any of the associated texts, revealing that the images can leak such sensitive social relationships, which may be against the intention of the uploader. Similarly, we can use the same approach to infer social relationships between different people appearing in an image, which will be further investigated in our future work.

During the manual inspection of the images posted by the 1,621 uploaders mentioned in Finding~8, we also noticed another potential finding, which will require more future study to confirm. 73 accounts did not use a real face as their profile image, however, we can confidently infer that the most appearing face in the image timeline of the account is the uploader's real face. This may be a privacy issue if the uploader indeed does not want to reveal their face publicly. Re-identifying such uploaders' real faces can also help better distinguish the uploader from other subjects and bystanders to facilitate the detection of other photographed persons in face images. We plan to explore this potential finding and its implications on face privacy in our future work.

\section{Further Discussions}
\label{sec:discussions}

In this section, we discuss applications, limitations, and future work of the three key contributions of our work.

\subsection{Practical Implications of Our Work}

\subsubsection{Deploying Bystander-subject Classifier in Real-world Cases}

Our proposed method to detect bystanders can effectively classify the subject and bystander in images. Our proposed semi-automated framework is based on this novel method to do the large-scale privacy measurement, which proved the usefulness of our classifier in the context of OSNs. Combined with image filters or encrypting technology, our classifier is easy to deploy in OSNs, image-sharing websites, and camera devices to automatically protect bystanders. Consider the following scenario where an OSN platform wants to deploy our bystander classifier to automatically check each uploaded image, show detected bystanders to the uploader, and ask them to confirm if such bystanders should be automatically anonymized. Such a feature can be easily incorporated into the existing image uploading pipeline of most OSN platforms, and asking the uploader to confirm bystander face anonymization can serve as a behavioral nudge to enhance uploaders' awareness of face privacy and to ultimately help protect more bystanders' privacy. Another application scenario could be the development of an uploader-facing web browser plug-in, which can be installed by an uploader without relying on the deployment of our classifier by the online platform. In this scenario, a privacy-aware uploader (e.g., a professional photographer) can leverage such a web browser plug-in to reduce their human efforts of anonymizing bystander faces before uploading images to multiple online platforms.

\subsubsection{Raising Awareness for Online Privacy}

The findings 
obtained via our large-scale analysis of the face privacy problem have profound operational implications for both online users and platforms. The data results reported in Section~\ref{subsec:Data_Results_of_Uploader_Posted_Faces} and Finding 2 show that we need to do more work to raise awareness among online users on face privacy, especially the privacy of bystanders. Finding 1 guides researchers to develop more useful privacy-preserving tools to help online users. Online users will benefit from being more informed about how to apply privacy protective measures more effectively. Online platforms should play a more active role in raising their users' awareness and deploying more user-centric tools for privacy protection. For example, the findings related to potential privacy breaches that we report in Section~\ref{subsec:Behaviors_Users_not-protect_faces} and Section~\ref{subsec:Leakage_Social_Attributes} can be used to warn people contained in images. The definitive findings we report in Section~\ref{subsec:Behaviors_Users_protected_faces} can be used to warn uploaders.

\subsection{Limitations and Future Work}

\subsubsection{Definition of Bystander}

As shown in Section~\ref{sec:overview}, how to define bystanders is a complex problem, and we adopted a practical definition for this work. In the future, we plan to conduct interviews with photographers, uploaders, and photographed people to gain deeper insights into detecting bystanders.

\subsubsection{Datasets}
Although our framework was applied to Twitter images only in this paper, our method is general enough to be applied to any images collected from other social media platforms. We plan to extend our work in the future to cover more platforms.

\subsubsection{Model Errors} 

Although our proposed bystander-subject classifier achieved a very good performance, there are still some limitations. One obvious constraint is that our classifier relies on the underlying face detection algorithm so its performance is naturally bounded. This is however not a limitation of our classifier. In addition, Dataset 1 we constructed may not have sufficient images for learning all useful features for classifying bystanders. In our future work, we plan to extend Dataset 1 with more images especially those with more subjects as we observed a lack of such images in our current Dataset 1. With more data, we can try more advanced classifiers, potentially getting rid of the dependency on the underlying face detection algorithm (e.g., by incorporating it seamlessly in the bystander-subject classifier).

Furthermore, in our study and in Darling's work~\cite{DarlingLL20}, we compared feature-based methods with methods that use cropped face regions as input for training deep neural networks (we employed MaskRCNN~\cite{he2017mask}, while Darling et al.~\cite{DarlingLL20} used CNN). Both studies demonstrated that feature-based models achieve higher accuracy compared to CNN-based models. This discrepancy may stem from the fact that training deep neural networks solely on face regions may overlook the broader contextual features of the entire image. This point is highlighted in our comparison with Darling et al.'s feature-based approach: while both approaches initially focused on facial features, our integration of additional local and global image-related features, such as the number of people and comparisons of face-to-image size, contributed to superior classification performance over Darling et al.'s classifier. Moving forward, we intend to explore more advanced deep neural networks and leverage other state-of-the-art algorithms such as transformers that take entire images as input and simultaneously handle face localization and classification tasks. Our current work with feature-based models establishes a foundational benchmark for future enhancements.

In addition, one limitation of our framework is that, like all machine learning based models, our subject-bystander classifier cannot achieve 100\% accuracy so there are always errors when applying it to large-scale measurements. However, All faces detected were manually inspected to remove any false positives so all errors of our bystander-subject classifier were actually checked and corrected. False negatives caused by the underlying face detector unfortunately could not be corrected because it was prohibitive to inspect all negative results given the large number of missed faces in some images, which is however a common limitation of any computational OSN analysis work based on large or big data. Given the high accuracy of the face detector we used, we do not think that missed face images can substantially change our findings. The framework also has dependencies on some core algorithms, especially those for manipulation detection. Further improving such algorithms will be important to reduce unnecessary human efforts.

\subsubsection{Single-modal Classifier and Semi-automated Framework}

Our proposed classifier considers an image as the only input for the following reasons. First, research by Hasan et al.~\cite{hasan2020bystander_privacy} indicated that visual features in photos can effectively capture active posing and the willingness of individuals photographed. This aligns with our definition of subjects and bystanders based on their active participation in the photo shoot. Thus, visual cues in the photos can effectively distinguish between subjects and bystanders. Second, integrating text as a new modality presents challenges in data annotation and dataset construction. Annotators would need to not only analyze image content but also read and interpret textual information to annotate data accurately. Moreover, some of the websites from which (e.g., Douban~\cite{Douban}) we sourced our dataset often provide only photos without accompanying textual descriptions, limiting the feasibility of using text information. Considering that most images posted by online uploaders are accompanied by some textual content (in the original post and replies) and the textual information can potentially provide useful information for classifying bystanders from subjects, e.g., the underlying scenario of the image, the photographer's and/or subjects' intention to take the image, social relationships of people in the image, it will be useful to add text-based features using natural language processing (NLP) techniques to construct a dual-modal classifier with a richer set of features. Our single-modal classifier could serve as a valuable tool for assisting in the annotation tasks of multimodal classifiers in the future.

The semi-automatic analysis tool we developed currently relies entirely on images for reasons similar to those mentioned previously: integrating tweet text for analysis can greatly increase the complexity and cost. However, adding text analysis will help us better determine whether any protective measures for faces are motivated by the privacy-related intent of the uploader. Our current image-based analysis provides a reliable upper bound for such behavior in the analyzed images. Additionally, incorporating text analysis will help distinguish different categories of individuals in the photo, such as the subject's and the bystander's true social relationship with the uploader, thereby increasing the granularity of the analysis. This multimodal analysis is planned for our future work.


Another limitation is that we considered only the profile image of the account owner. If we also consider profile images of the account owner's friends and even friends of friends (FoFs), i.e., adding social link analysis, we may be able to link other subjects in an image with confirmed friends of the account owner, therefore allowing us to infer more useful information about the corresponding face privacy scenario. In addition, similar to the case of the bystander-subject classifiers, we can also consider textual content associated with an image posted to infer more about other subjects in the image, e.g., to identify the accounts of some subjects. It is also important to note that some users opt to use someone else's photo as their profile picture. To gain a more comprehensive understanding, future analyses may require the integration of additional data to determine whether users are using their actual faces as profile images.

\subsubsection{Our Analysis Method}

We employed a mixed method, encompassing both quantitative and qualitative analyses. Many findings and the results are essentially quantitative based on descriptive or inferential statistics. The qualitative analyses include manual encoding of some images, which led to thematic codes that are important to support analyses of all findings. The categorization of findings is also a qualitative process. While also based on quantitative evidence, many findings are more based on qualitative analysis, e.g., Findings 1, and 4-8. Our measurement and privacy analysis is an objective factual measurement study in which only the anonymized behaviors implemented by uploaders are identified as relevant to privacy.

Of course, although our analysis is already quite comprehensive, there are still some limitations and areas for further work. For instance, some findings are based on a smaller number of images or faces, and some quantitative results can be better explained if we conduct an empirical study with recruited online users to get their views on the corresponding behavioral aspects. Some quantitative analysis can also be enhanced by more advanced algorithms, e.g., the scenario-based analysis can benefit from algorithms that can support more scenarios potentially using NLP-based analysis of textual content associated with the image analyzed. We can also utilize an image-based social relationship inference classifier~\cite{ZhangLLT15, SunSF17} to analyze the dynamics between individuals in the photos, their connections with the uploader, and the actual social relationships among bystanders and the uploader (whether they are strangers or have social ties but are not actively participating in the photo). In addition, the deterministic relationship between the uploaders' behaviors of (not) anonymizing faces and the actual face privacy leakage still requires more research on the uploader's intention as well as the photographed people's awareness, willingness, and permission, which will be part of our future work. In order to better understand the attitudes of users towards the leakage of face privacy on the OSNs platform, as part of our future work we plan to conduct an online survey and some interviews with online users to enrich the data we can use to draw insights about behaviors of uploaders and those who were photographed.


\section{Ethical Considerations}
\label{sec:Ethical_Considerations}

Our work did not directly involve recruitment of human participants, but the nature of our work required collection and analysis of images containing human faces (a special category of sensitive personal data). Note that all labeling work was either done by the authors or by a labeling company. Images in our four datasets were all collected from public sources: Dataset 1 from multiple public data sources; Dataset 2.B from the public image dataset COCO2017; and Datasets 2.A and 3 from Twitter using its public API. Since the images include sensitive personal data (human faces as biometric data) that we could not anonymize due to the nature of the work, we stored the collected data on a secure server of the first author's institution. All data collected were made accessible to the project team only via secure access control mechanisms.

For the data collection for Datasets 2.A and 3 from Twitter and the privacy analysis of Dataset 3, we received IRB approval for our study. For collecting and using non-OSN public images in Datasets 1 and 2.B, we followed the standard practice in AI-related research on collecting public data and did not go through a research ethics review.

Last but not least, in order to allow other researchers to reconstruct the datasets we used for reproducibility purposes and for conducting follow-up research, we released complete information about our research including instructions, relevant data and source code of the bystander classifier at \url{https://github.com/Yuqi-Niu/Bystander-Detection}.


\section{Conclusion}
\label{sec:conclusion}

This paper reports our work on a new machine learning-based bystander-subject classifier to support large-scale analysis of the face privacy problem. The bystander-subject classifier is trained on face-based features, and its performance exceeds that of the most recent state-of-the-art methods proposed by Hasan et al.~\cite{hasan2020bystander_privacy} and Darling et al.~\cite{Darling2019, DarlingLL20}, with a substantial margin for both OSN and non-OSN images. Based on the developed bystander-subject classifier, we introduced a semi-automated framework to facilitate quantitative and qualitative analysis of the face images at scale. Applying the framework to 27,800 Twitter images, we validated the practical usefulness of our bystander-subject classifier with eight key findings evidenced by quantitative and qualitative results, which revealed different aspects of Twitter users' behaviors regarding face privacy. The findings have practical implications for online users to be more privacy-aware, and also for online platforms to develop privacy protection tools. The researchers could benefit from our findings and future directions to advance the research in online image privacy, which is an under-researched area at the moment.

\section{Acknowledgments}

This work was partly supported by the National Key R\&D Program of China under the grant number 2023YFB3106501, funded by China's Ministry of Science and Technology. The first author's work was also partly funded by the China Scholarship Council (CSC).

\bibliographystyle{cas-model2-names}
\bibliography{main}

\appendix

\section{Additional Tables}
\label{appendix:tables}

Table~\ref{table:number_of_faces_for_subjects_bystanders} shows the data results of uploaders posted subjects and bystanders from the face level and  Table~\ref{table:Basic_statistics_of_user_posting_photos} shows the data results from the image and user level. We report the data results of uploaders anonymizing faces from the face, image, and user level in Table~\ref{table_Face-level_privacy_leakage/protection_statistics}, Table~\ref{table:photo-level_privacy_leakage+protection_statistics}, and Table~\ref{table:user-level_privacy_leakage+protection_statistics}, respectively. We report the number of faces related to face modification intentions in Tables~\ref{table:protected_bystander_face} and \ref{table:protected_subject_face}.

\begin{table*}[!htb]
\centering
\small
\caption{The number of faces of subjects and bystanders.}
\label{table:number_of_faces_for_subjects_bystanders}
\begin{tabularx}{\linewidth}{l XX XXc}
\toprule
& Subject & Bystander & Friend & Uploader & Bystander*\\
\midrule
All faces& 53,942 & 29,840 & 45,750 &  8,352 & 29,680\\
Unique faces & 38,081 & 28,507 & 35,577 & 2,585 & 28,363\\
\bottomrule
\end{tabularx}
\end{table*}

\begin{table}[!htb]
\centering
\caption{Basic data of uploader who posted images containing faces.}
\label{table:Basic_statistics_of_user_posting_photos}
\begin{tabularx}{\linewidth}{l YY}
\toprule
& Image Level & Uploader Level\\
\midrule
Only subject & 20,601 & 960\\
Only bystander & 150 & 1\\
Subject \& Bystander & 7,049 & 2,075\\
\midrule
Only friend & 14,849 & 680\\
Only uploader & 4,555 & 48\\
Only bystander* & 144 & 1\\
Friend \& Uploader & 1,271 & 240\\
Friend \& Bystander* & 5,551 & 1,126\\
Uploader \& Bystander* & 643 & 14\\
Friend \& Uploader \& Bystander* & 787 & 927\\
\bottomrule
\end{tabularx}
\end{table}

\begin{table}[!htb]
\centering
\caption{Face-level anonymization data.}
\label{table_Face-level_privacy_leakage/protection_statistics}
\begin{tabularx}{\linewidth}{lcc}
\toprule
 & Friend & Bystander*\\
\midrule
No anonymization & 45,184 & 29,574\\
Partial anonymization  & 78 & 21\\
Full anonymization & 488 & 85\\
\bottomrule
\multicolumn{3}{X}{\footnotesize Among those partially anonymized, the numbers of friends and bystanders modified for privacy purposes are 49 and 21, respectively.}
\end{tabularx}
\end{table}  

\begin{table*}[!htb]
\centering
\caption{Image- and uploader-level anonymization data. We use (a, b) to represent the case where (a) friend and (b) bystander* are included in an image. Each of the two variables (a and b) is a 3-bit integer, whose value includes 100, 010, 001, 110, 101, 011, and 111. The meanings of the three bits are as follows: the first bit -- a binary value (0 or 1) indicating if the image contains a non-anonymized face of a friend; the second bit -- a binary value indicating if the image contains a partially anonymized face of a friend; the third bit -- a binary value indicating if the image contains a fully anonymized face of a friend.}
\label{tab:privacy_leakage+protection_statistics}
\begin{subtable}[T]{0.47\linewidth}
\centering
\caption{Image-level anonymization data.}
\label{table:photo-level_privacy_leakage+protection_statistics}
\begin{tabularx}{\linewidth}{l YY}
\toprule
 & Only Friend & Friends \& Uploader\\
\midrule
(100,-) & 14,562 & 12,63\\
(110,-) & 1 & 2\\
(010,-) & 40 &\\
(101,-) & 51 & 3\\
(001,-) & 195 & 3\\
\midrule
 & Only Bystander* & Bystander* \& Uploader\\
\midrule
(-,100) & 643 & 138\\
(-,010) &  & 1\\
(-,101) &  & 1\\
(-,001) &  & 4\\
\midrule
 & Friend \& Bystander* & Friend \& Bystander* \& Uploader\\
\midrule
(100,100) & 5,458 & 781\\
(100,010) & 12 &\\
(100,001) & 9 &\\
(100,110) & 2 &\\
(100,101) & 28 &\\
(100,011) & 1 &\\
(010,100) & & 1\\
(010,010) & 2 &\\
(010,110) & 1 &\\
(010,111) & 1 &\\
(001,100) & 3 & 2\\
(001,001) & 22 & 1\\
(001,101) & 8 &\\
(101,100) &  & 1\\
(101,001) & 2 &\\
(101,101) & 1 & 1\\
(011,011) & 1 &\\
\bottomrule
\end{tabularx}
\end{subtable}
\hspace{1em}
\begin{subtable}[T]{0.47\linewidth}
\centering
\caption{Uploader-level anonymization data.}
\label{table:user-level_privacy_leakage+protection_statistics}
\begin{tabularx}{\linewidth}{l YY}
\toprule
 & Only Friend & Friends \& Uploader\\
\midrule
(100,-) & 630 & 235\\
(110,-) & 3 & 2\\
(111,-) & 5 &\\
(011,-) & 1 &\\
(101,-) & 24 & 3\\
(001,-) & 17 &\\
\midrule
 & Only Bystander* & Bystander* \& Uploader\\
\midrule
(-,100) & 14 &\\
(-,010) & & 1\\
\midrule
 & Friend \& Bystander* & Friend \& Bystander* \& Uploader\\
\midrule
(100,100) & 1,037 & 878\\
(100,110) & 1 &\\
(100,101) & 1 & 2\\
(001,100) & 2 & 1\\
(001,001) & 3 & 1\\
(110,100) & 7 & 7\\
(101,100) & 48 & 30\\
(101,001) & 7 & 1\\
(101,101) & 7 & 4\\
(011,101) & 1 &\\
(011,011) & 1 &\\
(111,100) & 3 & 3\\
(111,010) & 1 &\\
(111,001) & 1 &\\
(111,110) & 3 &\\
(111,101) & 3 &\\
\bottomrule
\end{tabularx}
\end{subtable}
\end{table*}

\begin{table*}[!htb]
\centering
\caption{The numbers of faces for different inferred intentions of the uploaders who manipulated faces of bystanders*. Empty cells indicate zero.}
\label{table:protected_bystander_face}
\begin{tabularx}{\linewidth}{l YYYY}
\toprule
Intention & No Anonymization & Partial Anonymization & Full Anonymization & Uploader\\
\midrule
Privacy &  & 20 & 79 &\\
Privacy \& Beauty & & & &\\
Privacy \& Humor & & & &\\
Privacy \& Information & & 1 & &\\
\midrule
Beauty & 7 & & 5 & 1\\
Beauty \& Information & & &\\
Humor & & & 1 &\\
Information & & & &\\
Unknown & & & &\\
\bottomrule
\end{tabularx}
\end{table*}

\begin{table*}[!htb]
\centering
\caption{The numbers of faces for different inferred intentions of the uploaders who manipulated faces of friends. Empty cells indicate zero.}
\label{table:protected_subject_face}
\begin{tabularx}{\linewidth}{l YYYY}
\toprule
Intention & No Anonymization & Partial Anonymization & Full Anonymization & Uploader\\
\midrule
Privacy & & 34 & 453 &\\
Privacy \& Beauty & & 14 & 11 &\\
Privacy \& Humor & & 1 & 11 &\\
Privacy \& Information & & & &\\
\midrule
Beauty & 54 & 14 & 8 & 15\\
Beauty \& Information & 3 & & &\\
Humor & 2 &  14 & 4 & 1\\
Information &  8 & & 1 & 3\\
Unknown & & 1 & &\\
\bottomrule
\end{tabularx}
\end{table*}

\section{Manipulation Intention}
\label{appendix:Manipulation Intention}

Our study involves inferring the motivations behind users modifying faces in photos, a process conducted by three authors. Initially, one author reviewed 50\% photos with modified faces and developed a codebook with five categories: privacy, beauty, humor, information, and unknown. Subsequently, the other two authors used this codebook to annotate the same 50\% images and discussed the codebook. The three annotators then independently coded all the modified faces based on the discussed codebook, after which the three annotators discussed any discrepancies in their annotations and reached a consensus. Images for which consensus could not be achieved were labeled as ``unknown''.

The annotators inferred the motivation of the uploader based on the overall context of each photo, including the scene, activities of the individuals, the location and method of face modification, the sentiment expressed from the photo, and the category of the person being modified (e.g., children are often modified for privacy reasons). Specific visual cues further guided the inference process:

\begin{itemize}
\item Beauty: Faces with common beauty filters and stickers, such as those found in Snapchat's beauty filters (\url{https://www.snapchat.com/explore/beauty/lenses}), indicating an intention to enhance attractiveness.

\item Information: Faces containing text, such as usernames or IDs, suggesting an informational purpose.

\item Humor: Faces with filters and effects designed for humor, such as face swaps, dog ears, or funny distortions (e.g., \url{https://www.snapchat.com/explore/funny/lenses}), or those resembling meme formats (e.g., \url{https://www.pinterest.com/digitalmomblog/funny-memes/}), indicating a humorous intent.

\item Privacy: Faces that are highly pixelated or blurred, making them unrecognizable, especially when the photo content is neither controversial nor humorous, indicating a focus on maintaining privacy.
\end{itemize}

\begin{figure*}[!htb]
\centering
\subcaptionbox{Original photo}{\includegraphics[width=\sfigwidth]{figures/No_anonymization.jpg}}
\hspace{1em}
\subcaptionbox{Beauty}{\includegraphics[width=\sfigwidth]{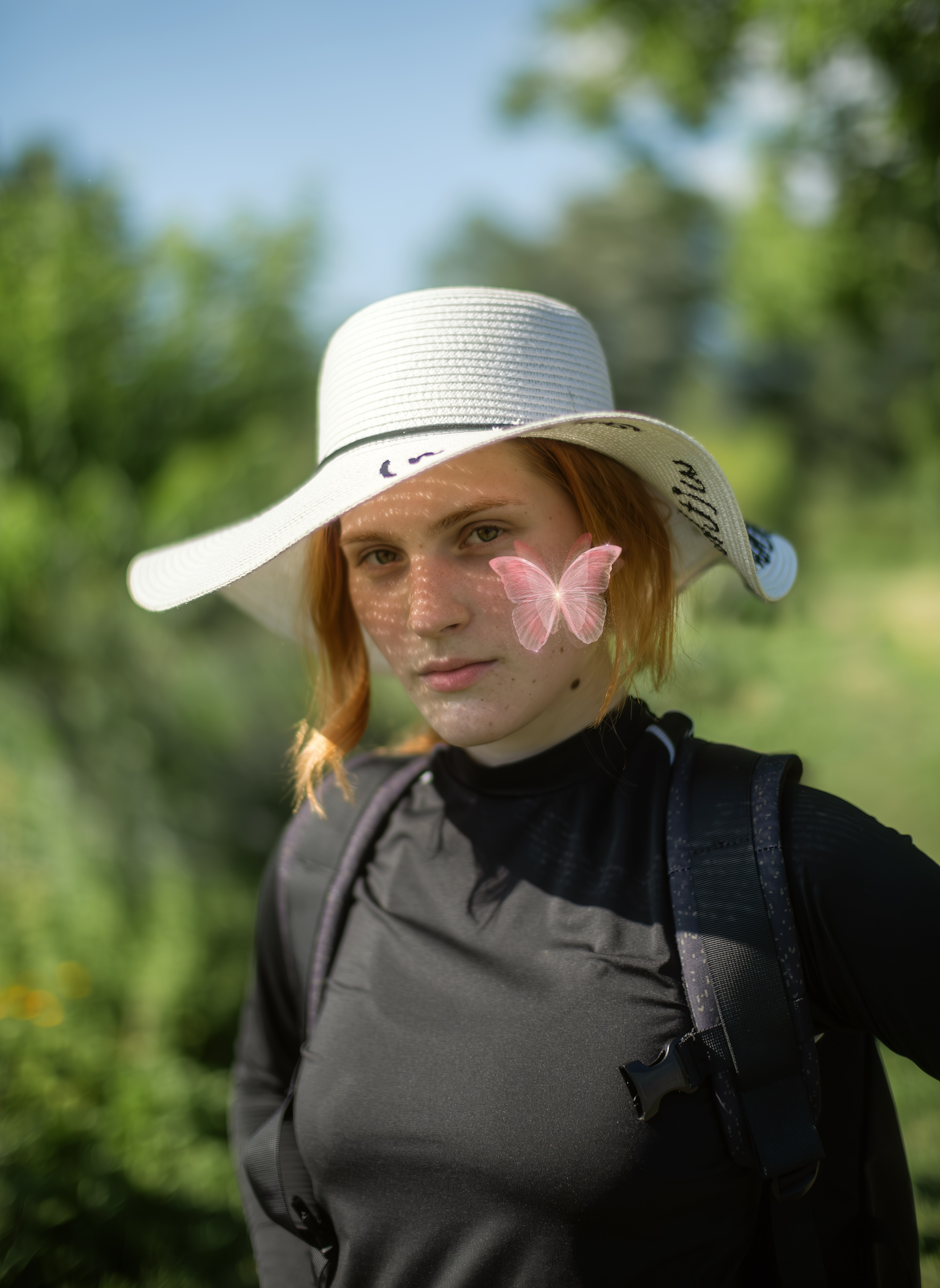}}
\hspace{1em}
\subcaptionbox{Information}{\includegraphics[width=\sfigwidth]{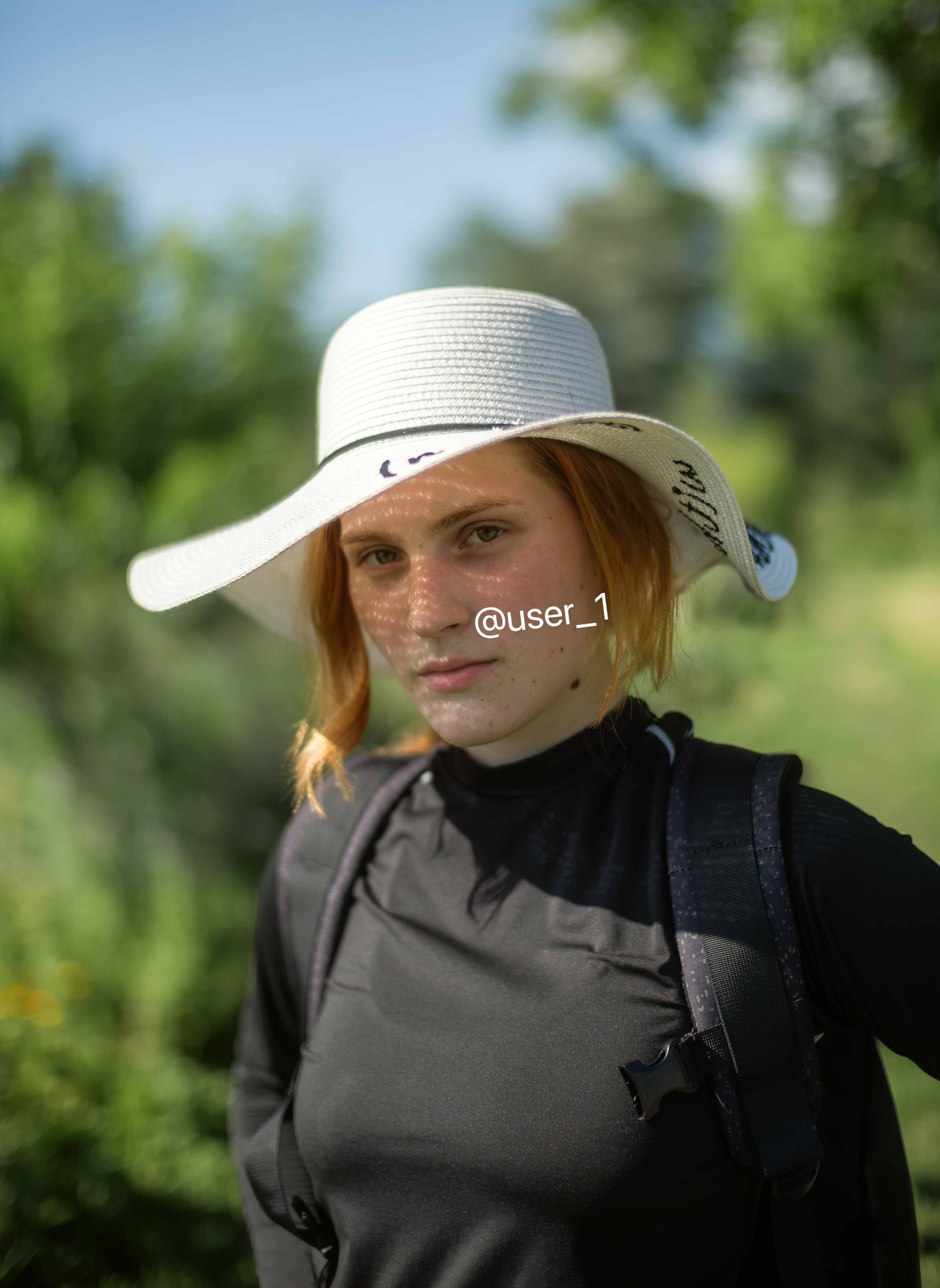}}
\hspace{1em}
\subcaptionbox{Humor}{\includegraphics[width=\sfigwidth]{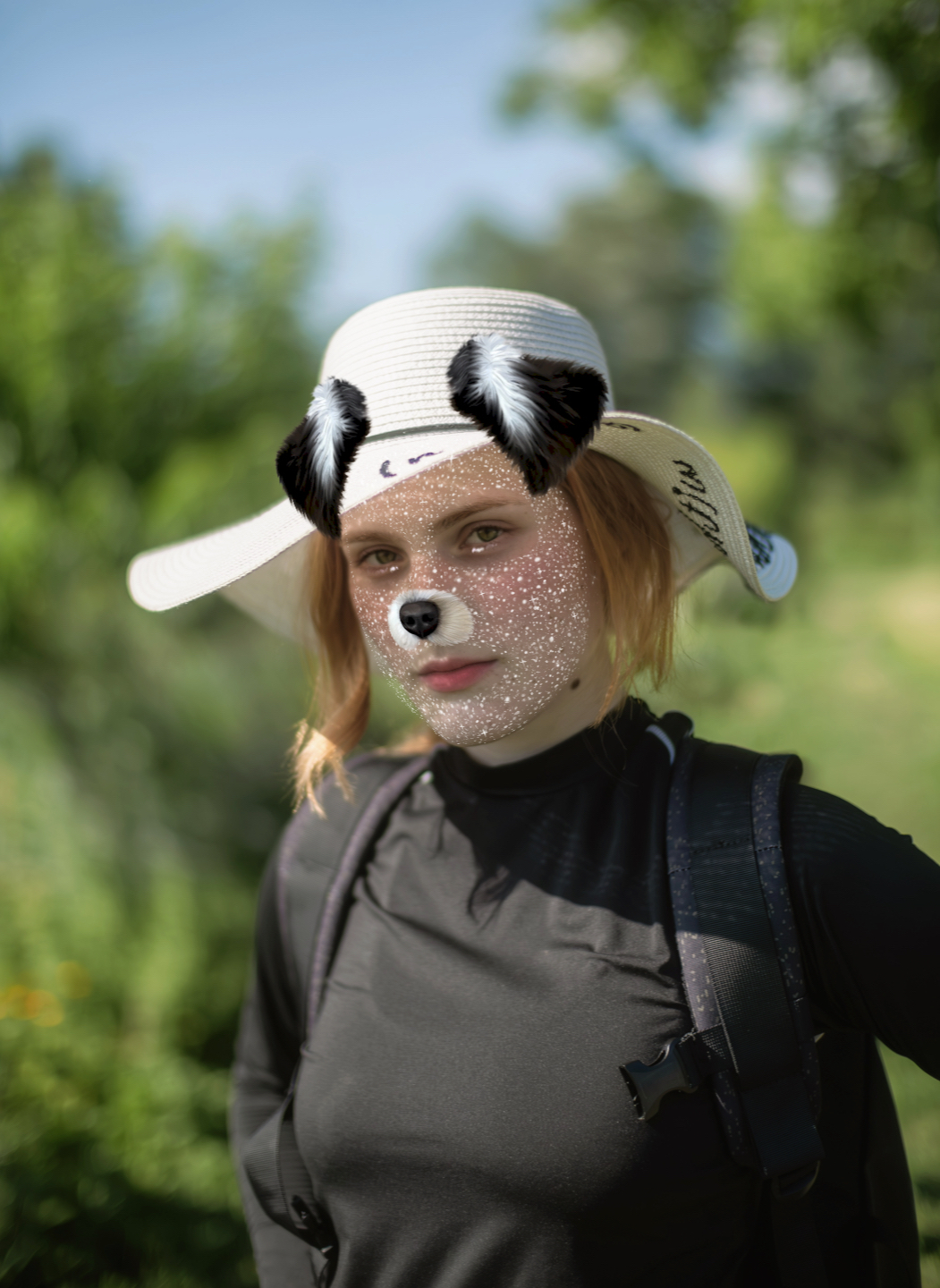}}
\hspace{1em}
\subcaptionbox{Privacy}{\includegraphics[width=\sfigwidth]{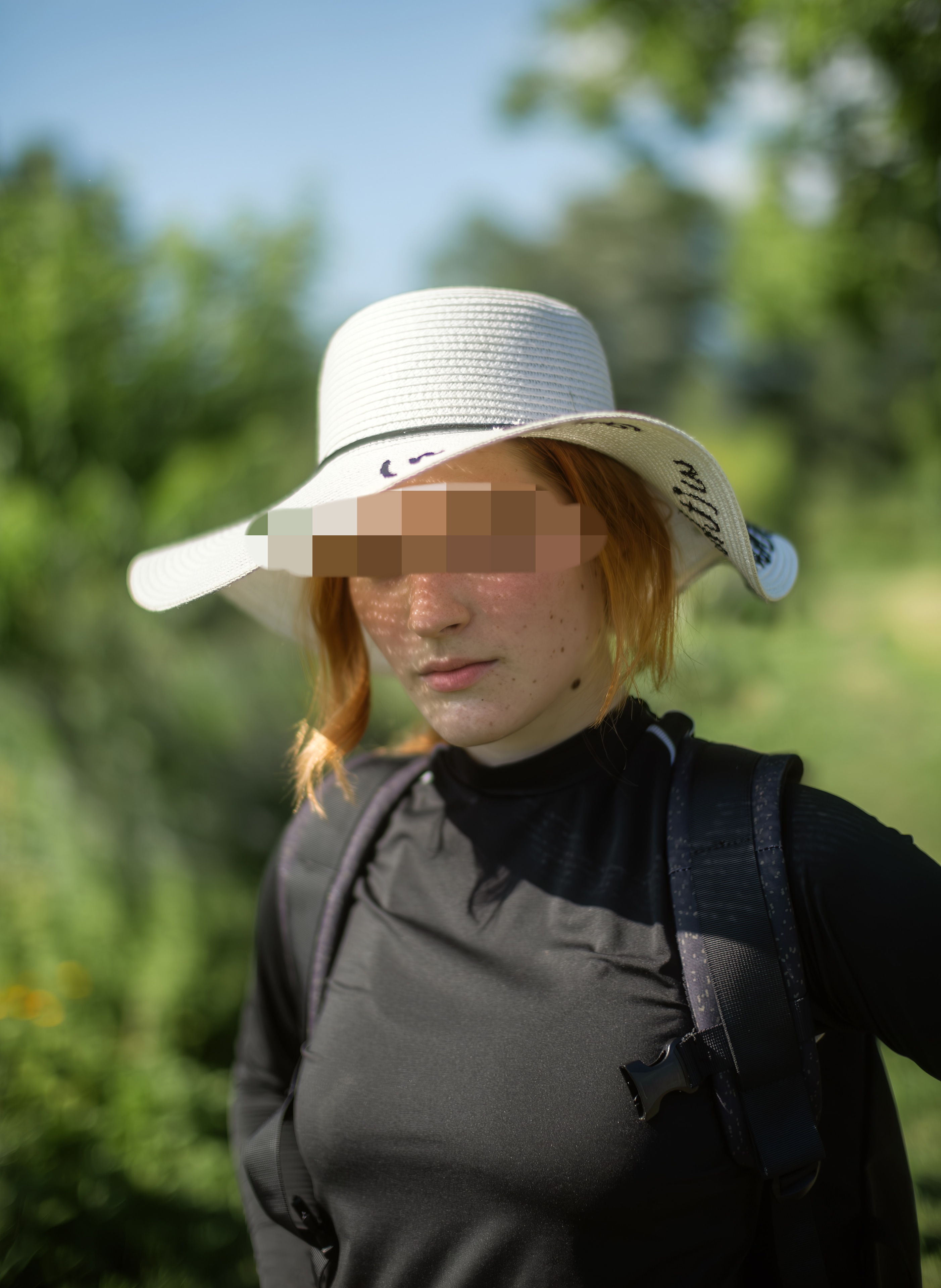}}
\caption{Examples of anonymized face with different Manipulation Intention.}
\label{figure:example1}
\end{figure*}

Based on the photo used in Figure~\ref{figure_example-anonymized-photos}, we have included the processing methods observed during our analysis to illustrate the inferred purposes behind these modifications (Figure~\ref{figure:example1}).

\end{document}